\documentclass{PoS}

\usepackage{epsfig}
\usepackage{graphicx}

\title{Nuclear Symmetries of the similarity renormalization
group for nuclear forces}

\ShortTitle{Symmetries of the similarity renormalization group}

\author{\speaker{E. Ruiz Arriola}%
         \thanks{E.R.A. was supported by Spanish DGI
  (grant FIS2011-24149) and Junta de Andaluc{\'{\i}a} (grant FQM225).
S.S. was supported by Instituto Presbiteriano Mackenzie through Fundo Mackenzie
de Pesquisa and FAPESP and V.S.T. by FAEPEX/PRP/UNICAMP and FAPESP.
Computational power provided by FAPESP grants 2011/18211-2 and 2010/50646-6.
}\\
Departamento de F\'{\i}sica At\'{o}mica, Molecular y 
  Nuclear and Instituto Carlos I de  F{\'\i}sica Te\'orica y Computacional. 
Universidad de Granada, E-18071 Granada, Spain.\\
        E-mail: \email{earriola@ugr.es}}

\author{V. S. Tim\'oteo\\
Faculdade de Tecnologia, Universidade Estadual de Campinas,
13484-332, Limeira - SP - Brazil \\
 E-mail: \email{varese@ft.unicamp.br}}

\author{S. Szpigel \\
Centro de Ci\^encias e Humanidades, Universidade Presbiteriana Mack
enzie,
01302-907, S\~ao Paulo - SP - Brazil \\ 
E-mail: \email{szpigel@mackenzie.br}}
\abstract{We review the role played by long-distance symmetries within
  the context of the similarity renormalization group (SRG)
  approach. This is based on phase-shift-preserving continuous unitary
  transformations that evolve Hamiltonians with a cutoff on energy
  differences. We find that there is a SRG cutoff, $\lambda_{\rm
    Wigner} \sim 3 {\rm fm}^{-1}$ for which almost perfect fulfillment
  of Wigner SU(4) symmetry is found at the two body level. This
  suggests to look for similar symmetry patterns for three- and
  four-body forces. We also analyze the impact of potentials based 
  on Chiral Perturbation Theory in Nuclear Structure calculations.}

\FullConference{The 7th International Workshop on Chiral Dynamics,\\
		August 6 -10, 2012\\
		Jefferson Lab, Newport News, Virginia, USA}

\begin{document}

\section{Introduction}

The use of effective interactions in Nuclear Physics has been the
traditional procedure to side-step the short-distance correlations
triggered by the hard core of the NN potential below relative
distances of about half a fermi~\cite{Coraggio:2008in,Bogner:2009bt}. While this is an acceptable
requirement, we take the viewpoint that any sensible definition of an
effective interaction should also unveil hidden symmetries so that
they become manifest and can be exploited in the solution of the
Nuclear Many Body Problem.

One outstanding and time honoured symmetry was suggested by Wigner in
1937~\cite{Wigner:1936dx} to study nuclear specroscopy and the
corresponding SU(4) spin-flavour symmetry group is generated by the
Lie algebra of isospin $T^a$, spin $S^i$ and Gamow-Teller $G^{ia}$
generators in terms of the one particle spin $\sigma_A^i$ and isospin
$\tau_A^a$ Pauli matrices,
\begin{eqnarray} 
T^a =\frac12 \sum_A\tau_A^a \, , \quad S^i &=& \frac12 \sum_A
\sigma_A^i \, , \quad  G^{ia} = \frac12 \sum_A \sigma_A^i \tau_A^a \, . 
\end{eqnarray} 
The one-nucleon irreducible representations is a quartet made of
a spin and isospin doublet $ {\bf 4}= (p\uparrow, p\downarrow,
n\uparrow, n\downarrow) =(S=1/2,T=1/2) $. NN states with
relative angular momentum $L$ and total spin $S$ and isospin $T$
fulfilling $(-1)^{S+L+T}=-1$ due to Fermi statistics correspond to an
antisymmetric sextet and a symmetric decuplet which, in terms of
$(S,T)$ representations of the $SU_S(2) \otimes SU_T(2)$ subgroup, are
\begin{eqnarray} 
{\bf 6}_A &=& (1,0) \oplus(1,0) \quad L=0,2, \dots \quad \to (^1S_0,
^3S_1), (^1D_2, ^3D_{1,2,3}) , (^1G_2, ^3G_{1,2,3}) , \dots \\ {\bf
  10}_S &=& (0,0) \oplus(1,1) \quad L=1,3, \dots \quad \to
\phantom{(^1S_0, ^3S_1),} \, (^1P_1, ^3P_{0,1,2}), (^1F_1, ^3F_{0,1,2}),
\dots
\end{eqnarray} 
In particular, one obtains $V_{^3S_1} (r) = V_{^1S_0}(r)$ which seems
verified for $r > 2 {\rm fm}$ (but not below). An amazing result is
the large $N_c$ justification of this symmetry to ${\cal O} (1/N_c^2)$
accuracy~\cite{Kaplan:1995yg,Kaplan:1996rk} which strongly suggests to
understand {\it in what sense} can the symmetry be ckecked in the much
studied NN interaction, as this is a direct consequence of the
underlying QCD dynamics.

A long distance interpretation of the symmetry has been given within a
large $N_c$ spirit recently, particularly the role in higher partial
waves and the companion Serber
symmetry~\cite{CalleCordon:2008cz,CalleCordon:2009ps,RuizArriola:2009bg}.
Within a Wilsonian approach saturation of effective parameters has
been observed in \cite{Arriola:2010hj} and in
\cite{NavarroPerez:2012qr} by two different methods. On a more
fundamental level, recent lattice calculations have observed Wigner's
symmetry at the potential level~\cite{Ishii:2006ec} and also at the
scattering length level for the unphysical pion masses about four
times larger than in the real world~\cite{Beane:2013br}. In the
present contribution we summarize the findings of our renalysis 
~\cite{Timoteo:2011tt}  based  on the Similarity Renormalization
Group and provide some outlook.

\section{Wigner Symmetry and Potentials}

\begin{figure}
\begin{center}
\includegraphics[height=4cm,width=3.5cm]{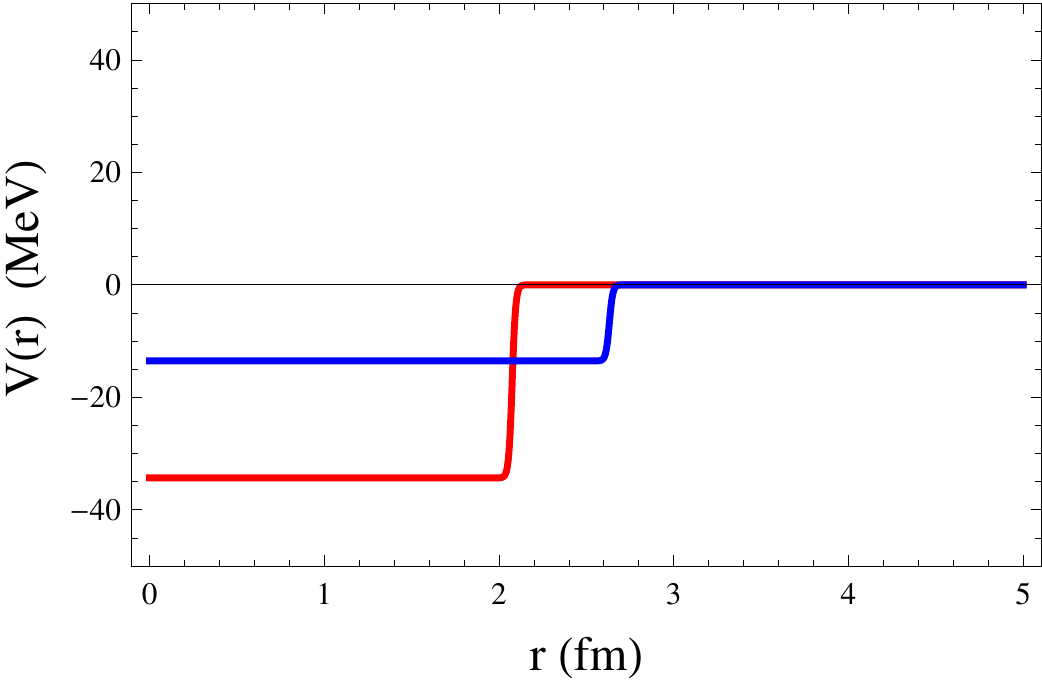}  
\includegraphics[height=4cm,width=3.5cm]{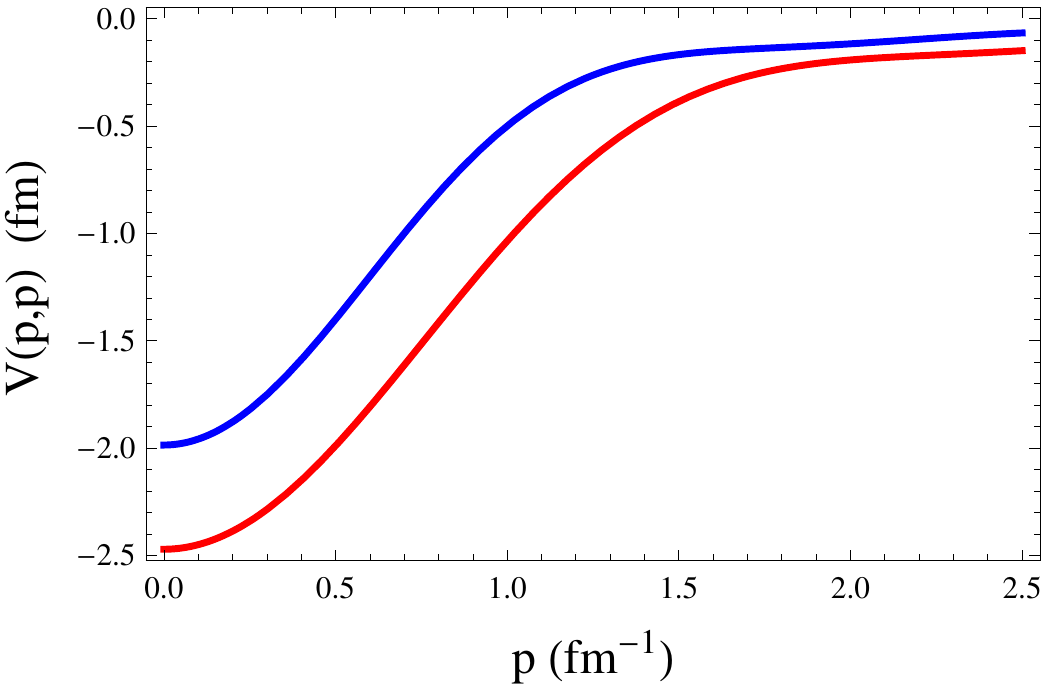}  
\includegraphics[height=4cm,width=3.5cm]{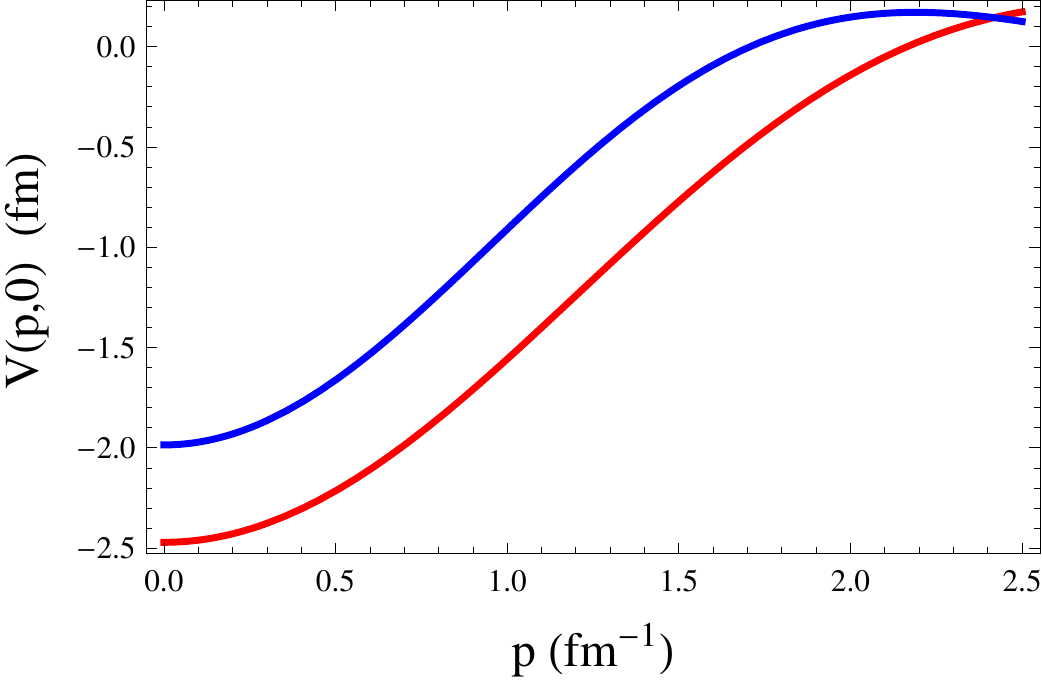} 
\includegraphics[height=4cm,width=3.5cm]{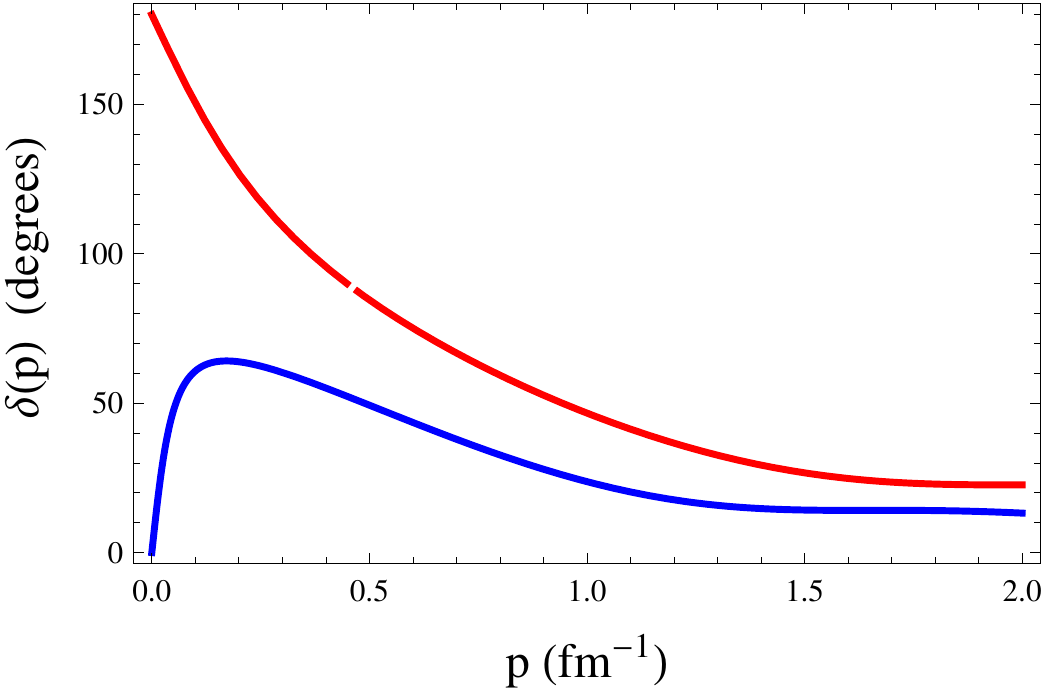} 
\end{center}
\caption{Square wells potentials, diagonal matrix elements,$ V(p,p)$,
  off-diagonal matrix elements, $V(0,p)$, and phase shifts for the
  $^1S_0$ (blue) and $^3S_1$ (red) channels. Parameters are adjusted
  to reproduce the scattering lengths and effective ranges (see main
  text).}
\label{fig:wigner}
\end{figure}

A rather simple way to see how Wigner symmetry emerges from low energy
NN-scattering data is by taking as an effective interaction a square
well potential of depth $-V_0$ and range $r_c$. The potential
parameters will be fixed by the corresponding scattering length
$\alpha_0$ and effective range $r_0$ given by the equations,
\begin{eqnarray}
\alpha_0 = r_c - \frac{\tan \sqrt{M V_0} r_c}{\sqrt{M V_0}} \, ,
\qquad r_0 = r_c \left[ 1- \frac{1}{\alpha_0 r_c M V_0} -
  \frac{r_c^2}{3\alpha_0^2} \right] \, .
\end{eqnarray}
We will, in addition, define the volume integrals of the potential as 
\begin{eqnarray} 
C_0 = \int d^3 x \, V_{\rm eff}(\vec x) = - \frac{4\pi}{3} V_0 r_c^3
\, , \qquad  C_2 = - \frac16 \int d^3 x \, r^2 \, V_{\rm eff}(\vec x) =- \frac{4\pi}{30} V_0 r_c^5
\end{eqnarray}
Now fixing the values in the $^1S_0$ ($\alpha_0 =-23.74 {\rm fm}$,
$r_0 =2.75 {\rm fm}$) and $^3S_1$ ($\alpha_0 =5.40 {\rm fm}$,
$r_0 =1.75 {\rm fm}$)channels, and imposing that there are no  $^1S_0$ bound states and just one $^3S_1$ bound state (the deuteron) we get $(C_{0,^1S_0},C_{0,^3S_1}) = (-1030, -1288) {\rm MeV} {\rm fm}^3 $ and 
$(C_{2,^1S_0},C_{2,^3S_1}) = (715 , 554){\rm MeV} {\rm fm}^5 $. 
%\begin{eqnarray}
%&^1S_0& \, , \quad  r_c=2.63 {\rm fm} \, , \quad V_0 = 13.49 {\rm MeV} \, , \quad C_0 = -1030 {\rm MeV} {\rm fm}^3 \, , \quad C_2 = 715 {\rm MeV} {\rm fm}^5 \\
%&^3S_1& \, , \quad   r_c=2.08 {\rm fm} \, , \quad V_0 = 34.22 {\rm MeV} \, , \quad C_0 = -1288 {\rm MeV} {\rm fm}^3 \, , \quad C_2 = 554 {\rm MeV} {\rm fm}^5  
%\end{eqnarray}
The numerical value for $C_0$ agrees with the EFT
estimate~\cite{Mehen:1999qs}.  The role of tensor force and higher
partial waves has been further explored~\cite{Arriola:2010hj}.  Note
that the volume integrals as well as the momentum space potential generically defined as 
\begin{eqnarray}
V_{l',l}^{JS}(p',p) = M_N \int_0^\infty j_{l'}(p'r) j_{l}(pr) V_{l'l}^{JS}(r) r^2
\end{eqnarray}
are very similar numerically (see Fig.~\ref{fig:wigner}), $V_{^3S_1}
\sim V_{^1S_0}$. For the phase-shifts the previous approximation is
crude but it makes sense for wavelengths larger than the range of the
NN interaction when compared to high quality
potentials~\cite{Stoks:1994wp}.

\begin{figure}
\begin{center}
\includegraphics[height=4cm,width=3.6cm]{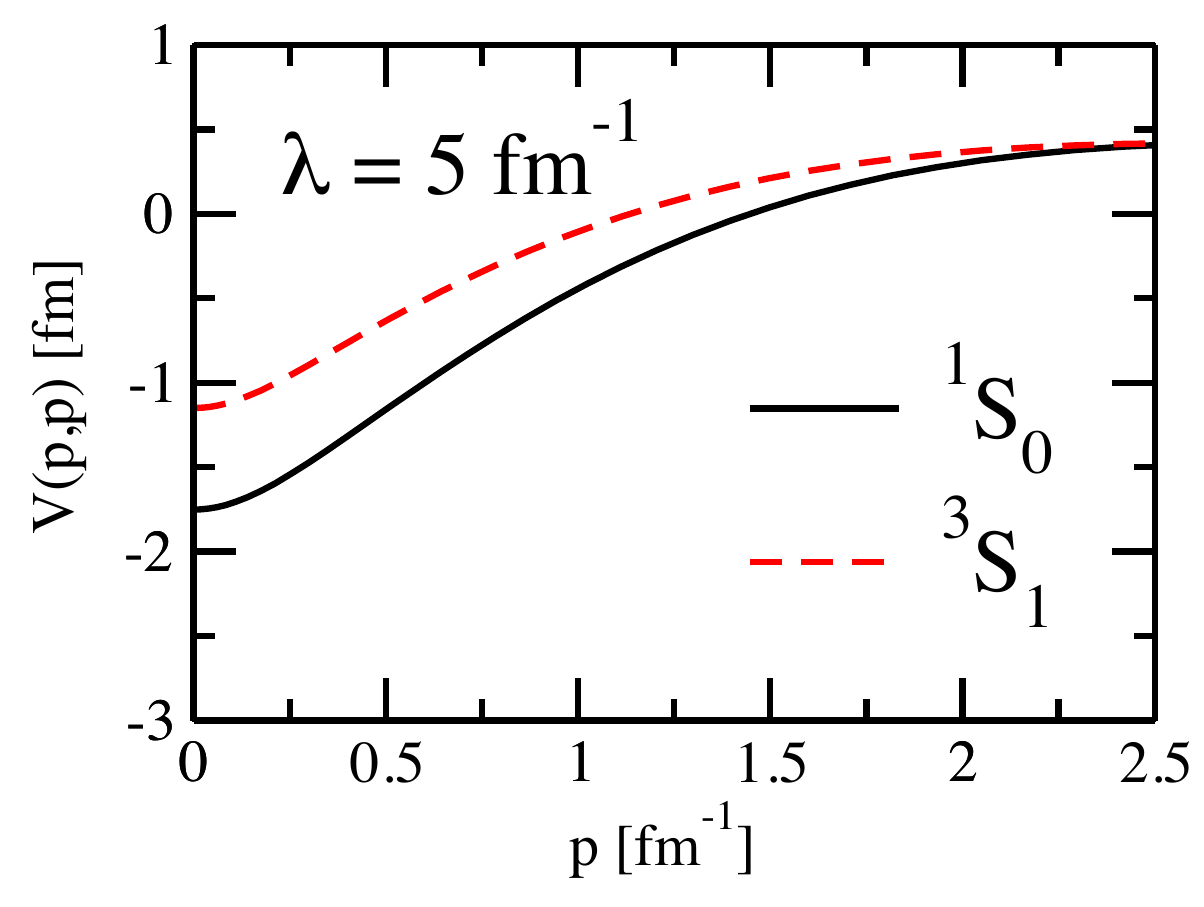}
\includegraphics[height=4cm,width=3.6cm]{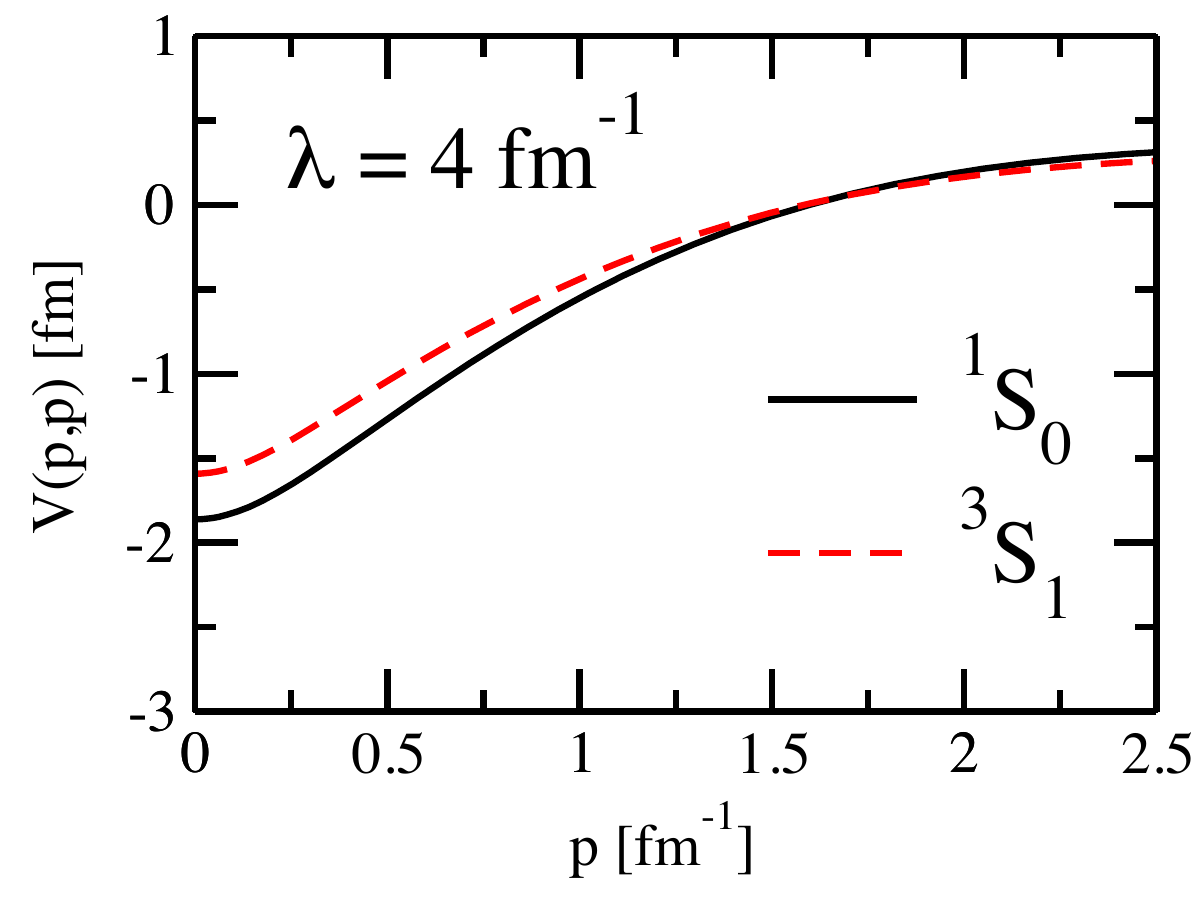} 
\includegraphics[height=4cm,width=3.6cm]{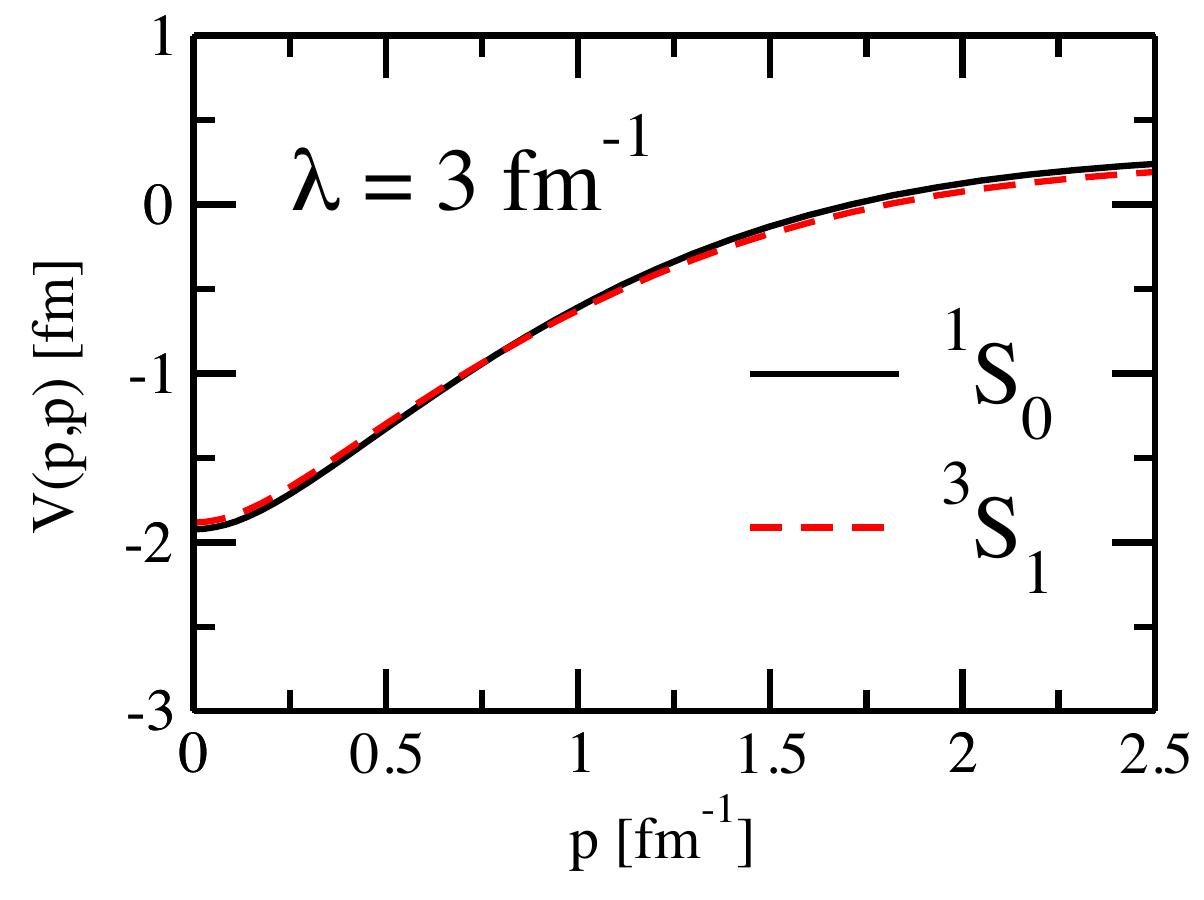} 
\includegraphics[height=4cm,width=3.6cm]{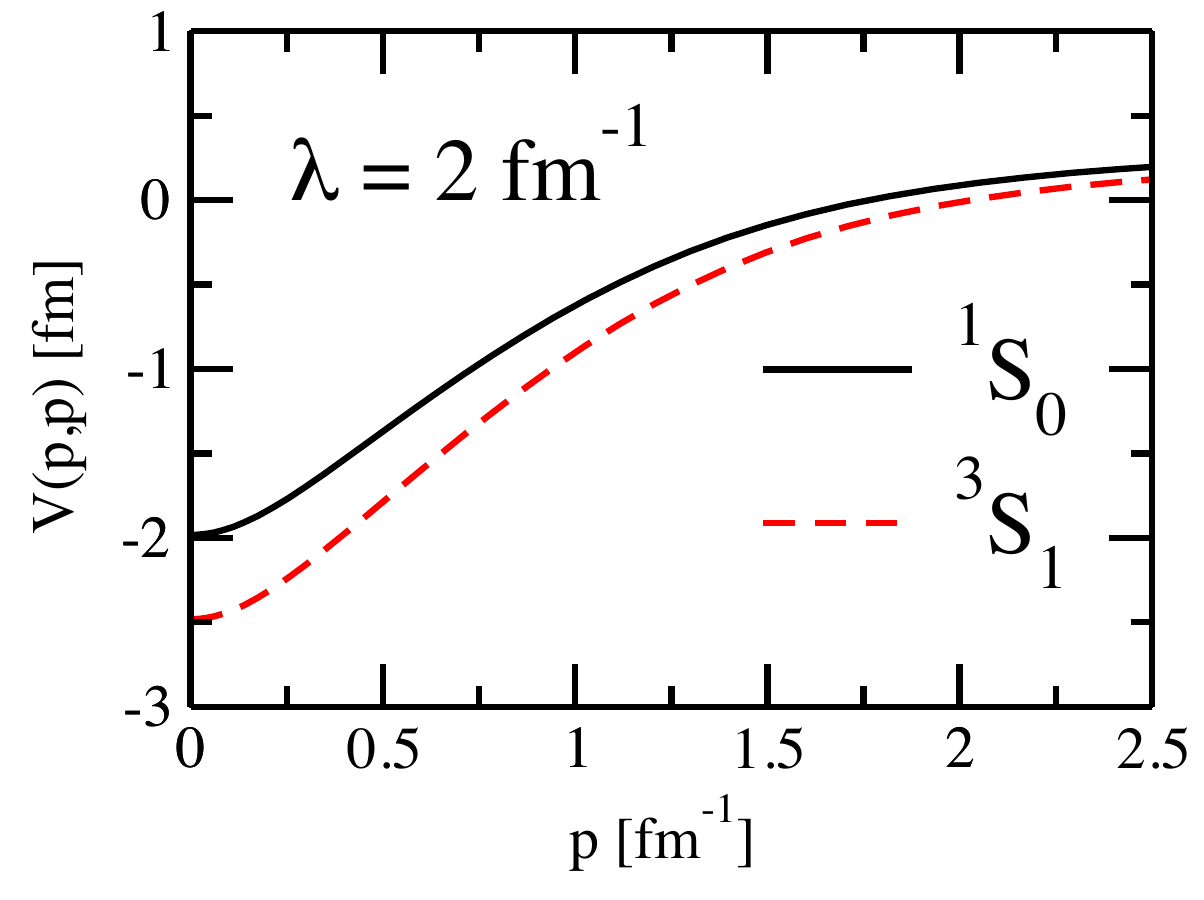}   \\
\includegraphics[height=4cm,width=3.6cm]{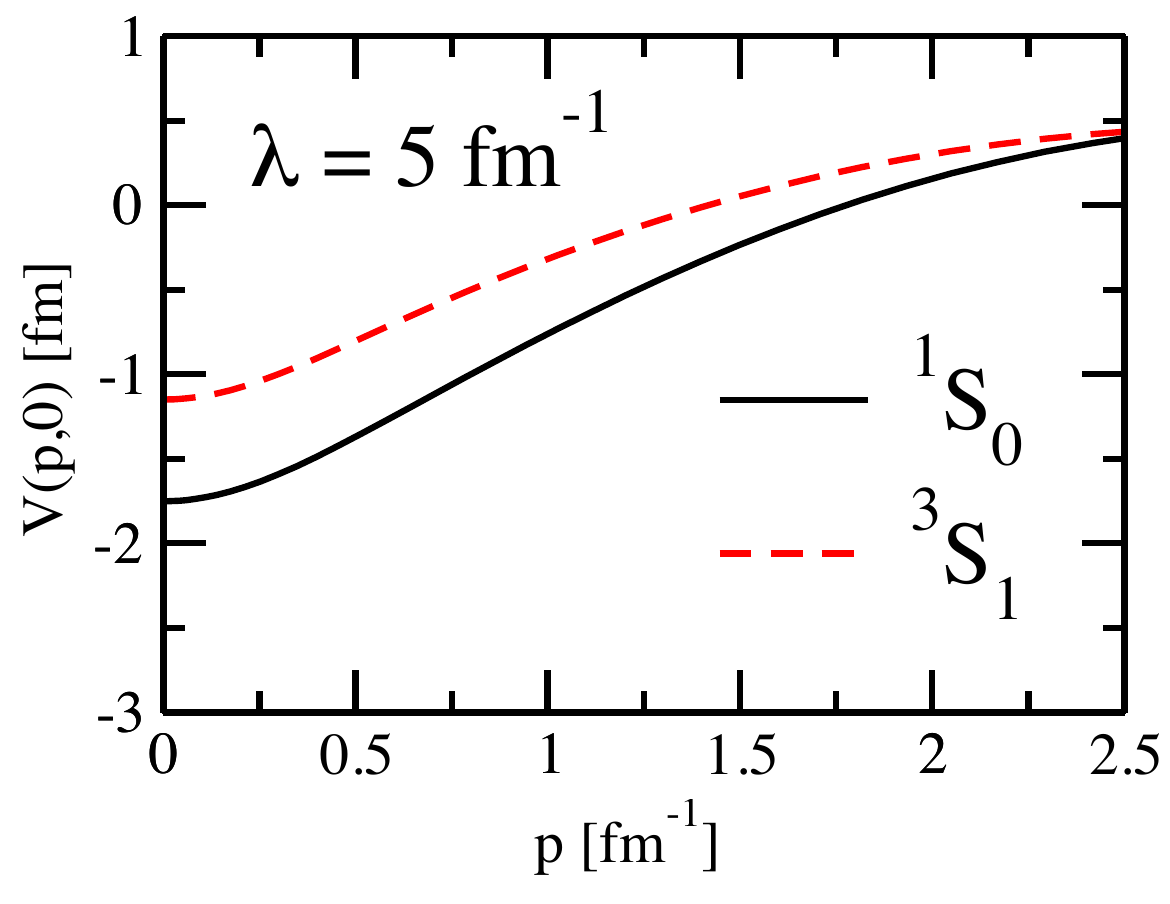}
\includegraphics[height=4cm,width=3.6cm]{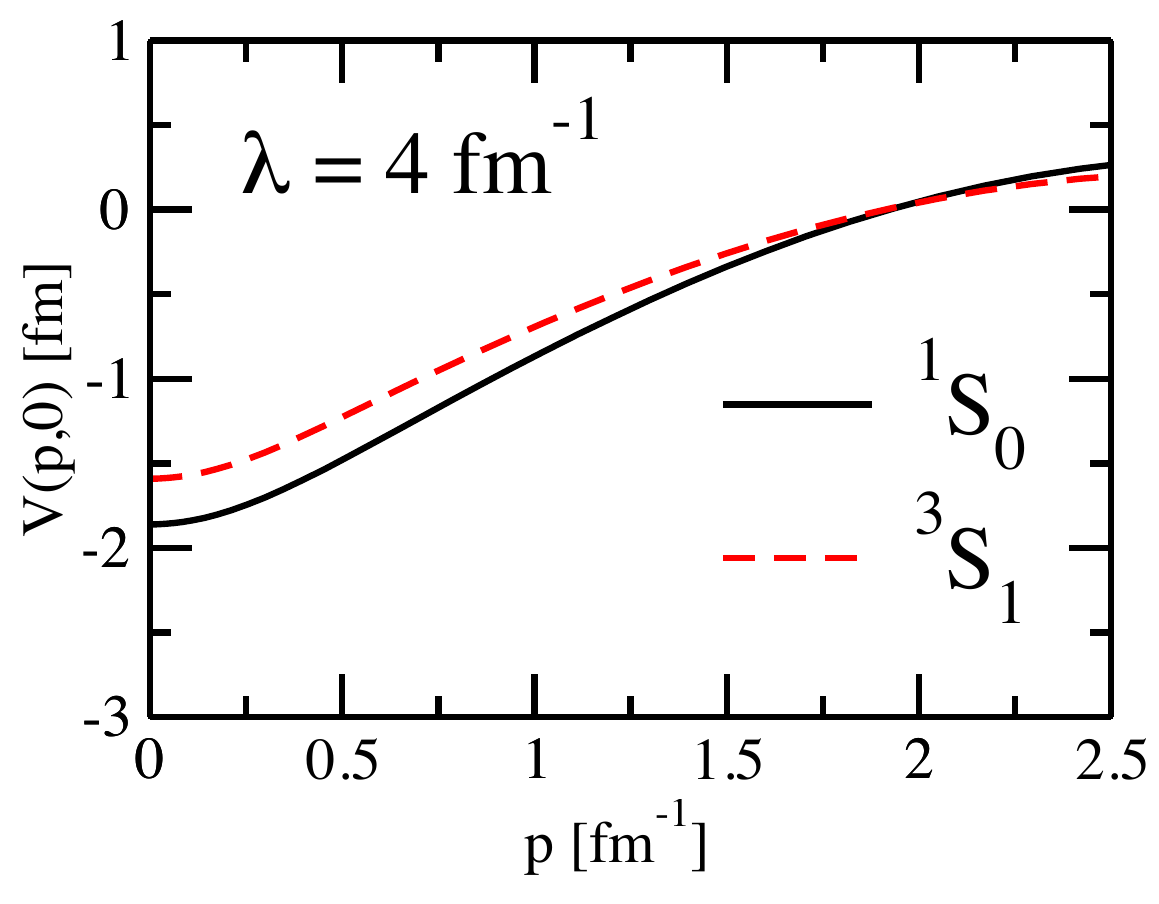}  
\includegraphics[height=4cm,width=3.6cm]{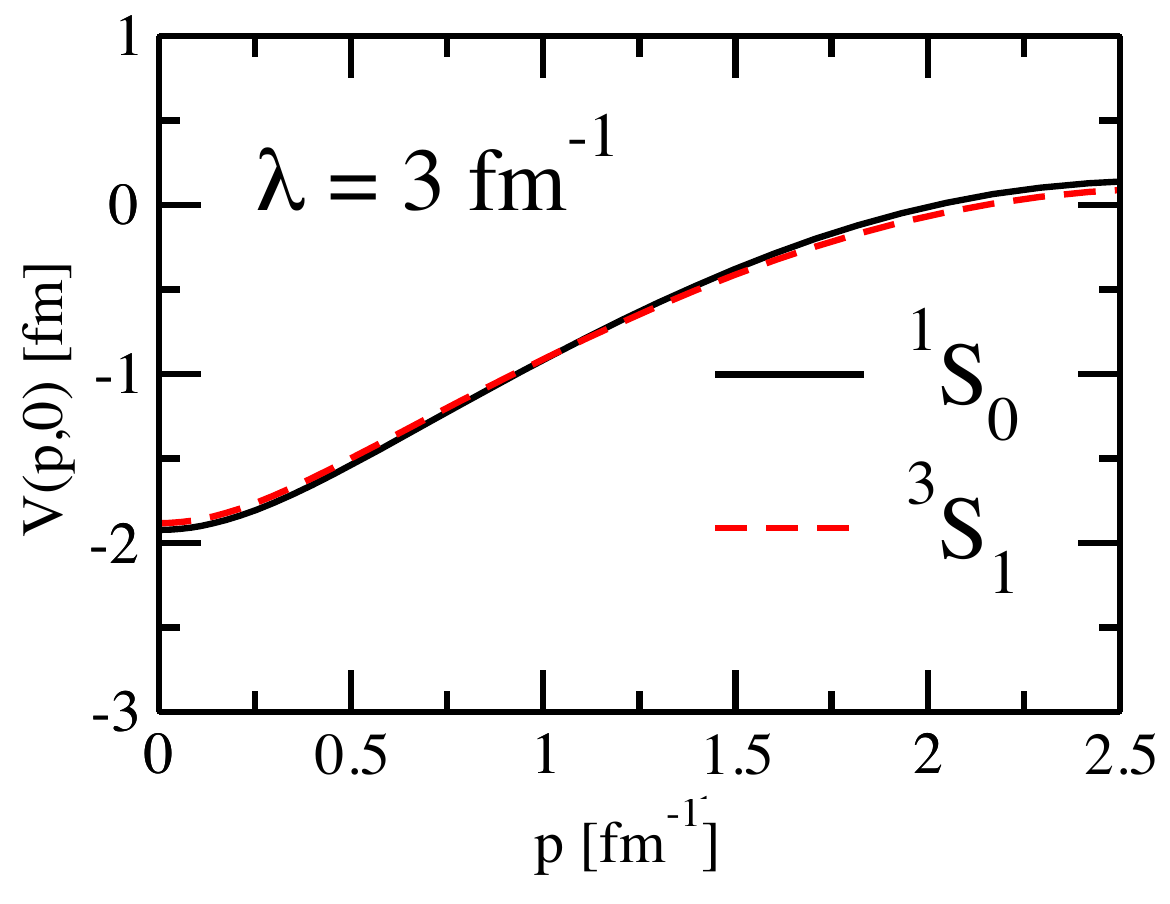}
\includegraphics[height=4cm,width=3.6cm]{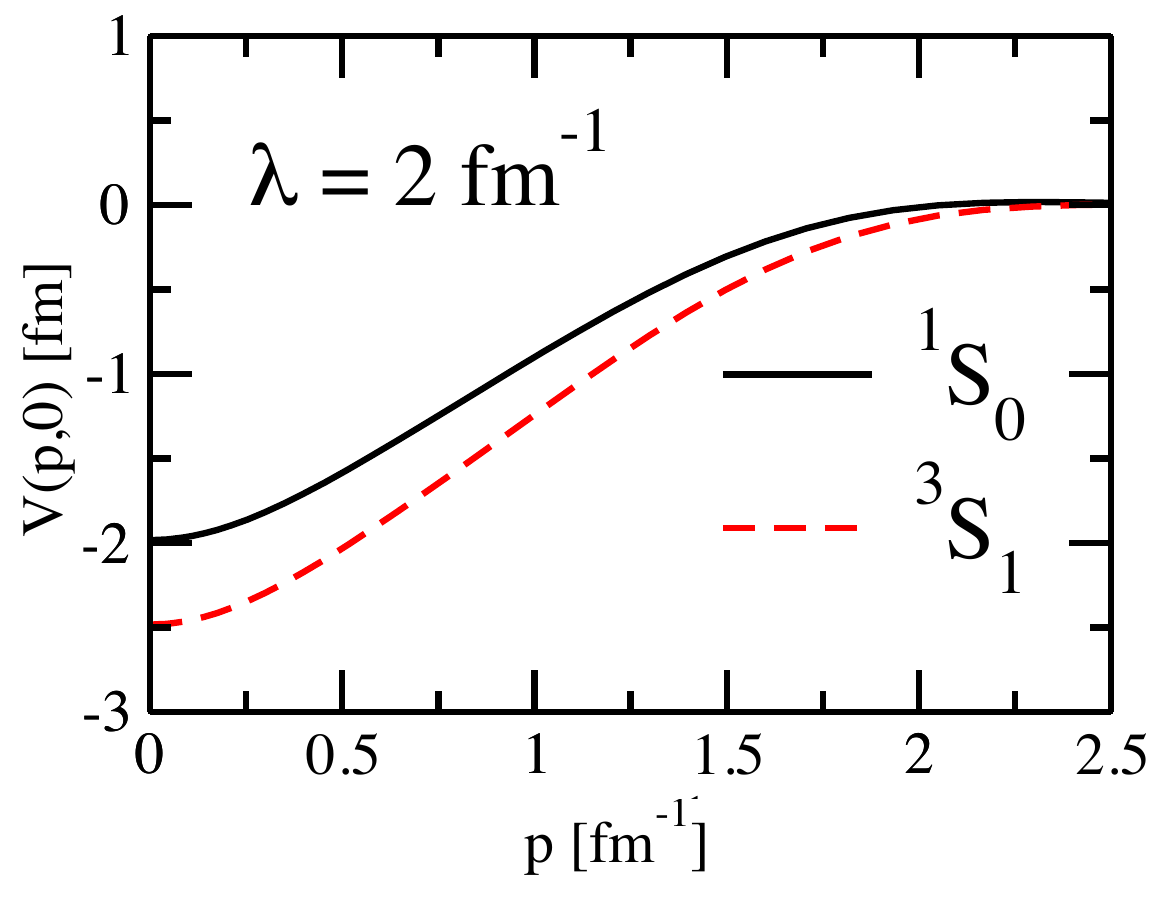}
\end{center}
\caption{Comparison between diagonal, $V(p,p)$, and fully
  off-diagonal, $V(p,0)$, matrix-elements of the SRG-evolved
  potentials for the $S$-waves (in {\rm fm}) as a function of the CM
  momentum $p$ (in ${\rm fm}^{-1}$), showing Wigner symmetry 
  for $\lambda_{\rm Wigner} \approx 3~ {\rm
    fm}^{-1}$.  We use the Argonne AV18 potential as the initial
  condition~\cite{Wiringa:1994wb}.}
\label{fig:AV18-wigner-S}
\end{figure}

\section{Wigner Symmetry and the SRG}

The previous analyses of the effective interaction and Wigner symmetry
are based on very low energy data. There is another rather surprising
way of unveiling the Wigner symmetry beyond this restricted range and
can be seen by using data up to about the lowest pion production
inelastic threshold. The SRG method has been amply used for NN
interactions in the last years~(\cite{Bogner:2006pc,Szpigel:2010bj})
and is based an an integro-differential equation for every (coupled) partial waves 
\begin{eqnarray}
&& -\frac14 \lambda^5 \frac{d V_\lambda (p',p) }{d\lambda} = -(p^2-p'^2)^2 V_\lambda (p',p) + \frac2{\pi}\int_0^\infty q^2 dq  \left( p^2+p'^2 - 2 q^2 \right) V_\lambda (p',q)V_\lambda (q,p) \, .
\end{eqnarray}
where $\lambda$ is the SRG cut-off. The solution generates from an
initial potential $V_{\lambda=\infty}(p',p)$ a one-parameter family of
phase-equivalent potentials at {\it all energies}, $\delta_\lambda (p)
= \delta_{\lambda=\infty}(p)$ which are driven to a stable fixed point
at $\lambda \to 0$~\cite{Timoteo:2011tt}. Large momentum-differences $|p-p'| \gg \lambda $ are suppressed as  
\begin{eqnarray}
V_\lambda (p',p) \approx V_{\lambda=\infty} (p',p) e^{-(p^2-p'^2)^2 /\lambda^4} + \dots 
\end{eqnarray}
implying a simplification of the nuclear many body problem. We show in
Fig.~\ref{fig:AV18-wigner-S} (compare with Fig.~\ref{fig:wigner}) the
results for the $^1S_0$ and $^3S_1$ channels for several SRG-cut-offs
for the AV18 potential~\cite{Wiringa:1994wb}. As we can see, there is
a scale $\lambda_{\rm Wigner} \sim 3 {\rm fm}^{-1}$ where
\begin{eqnarray}
V_{^1S_0,\lambda_{\rm Wigner}} (p',p) \approx V_{^3S_1,\lambda_{\rm Wigner}} (p',p) 
\end{eqnarray}
Of course, since the SRG transformation is unitary, the phase-shifts
remain the same for any value of $\lambda$ and despite the $^1S_0$ and
$^3S_1$ phase-shifts being {\it very different} the SRG-evolved and
phase-equivalent potentials look {\it very similar}.

\begin{figure}[pt]
\begin{center}
\includegraphics[height=3.4cm,width=3.5cm]{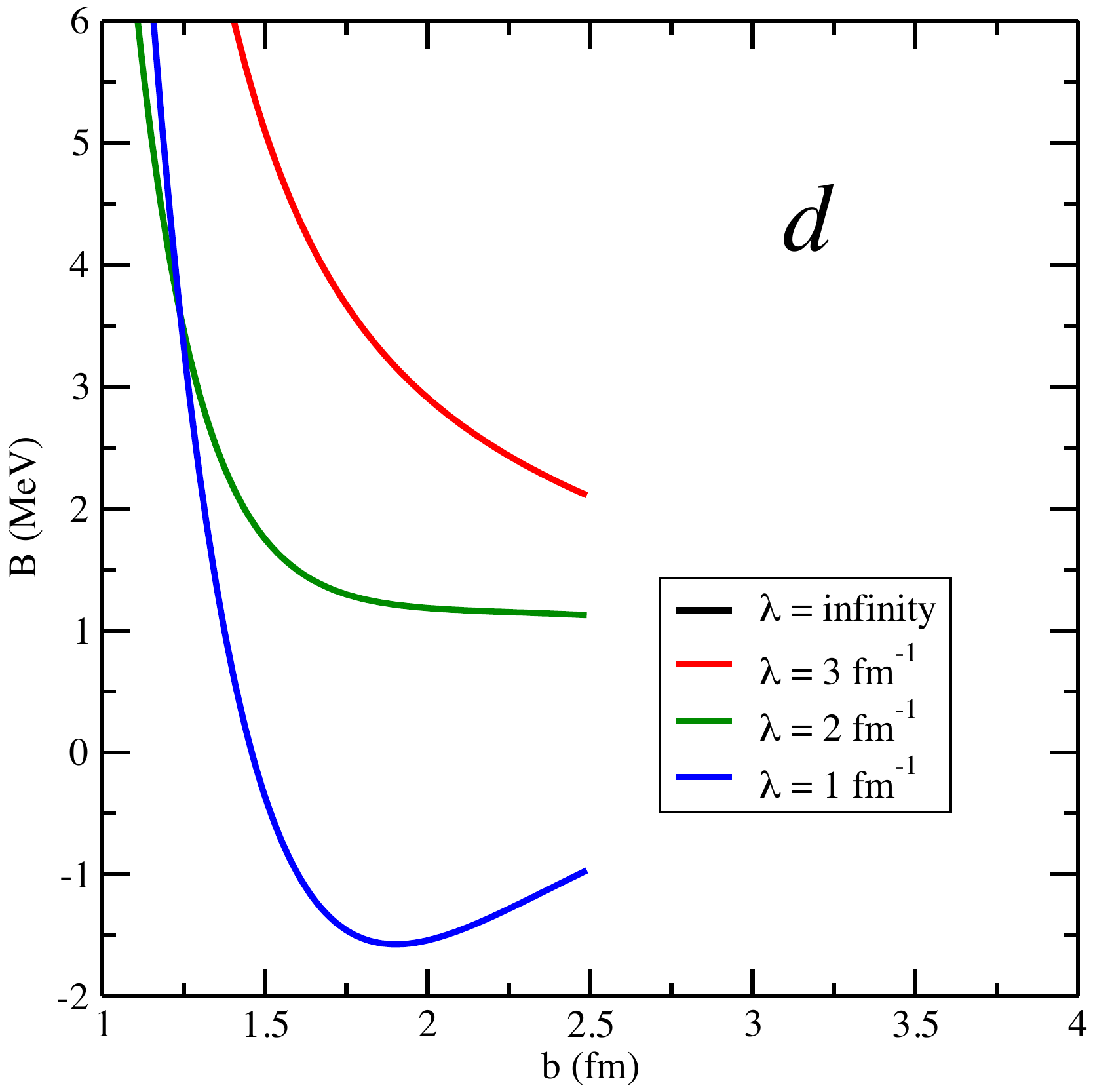}
\includegraphics[height=3.4cm,width=3.5cm]{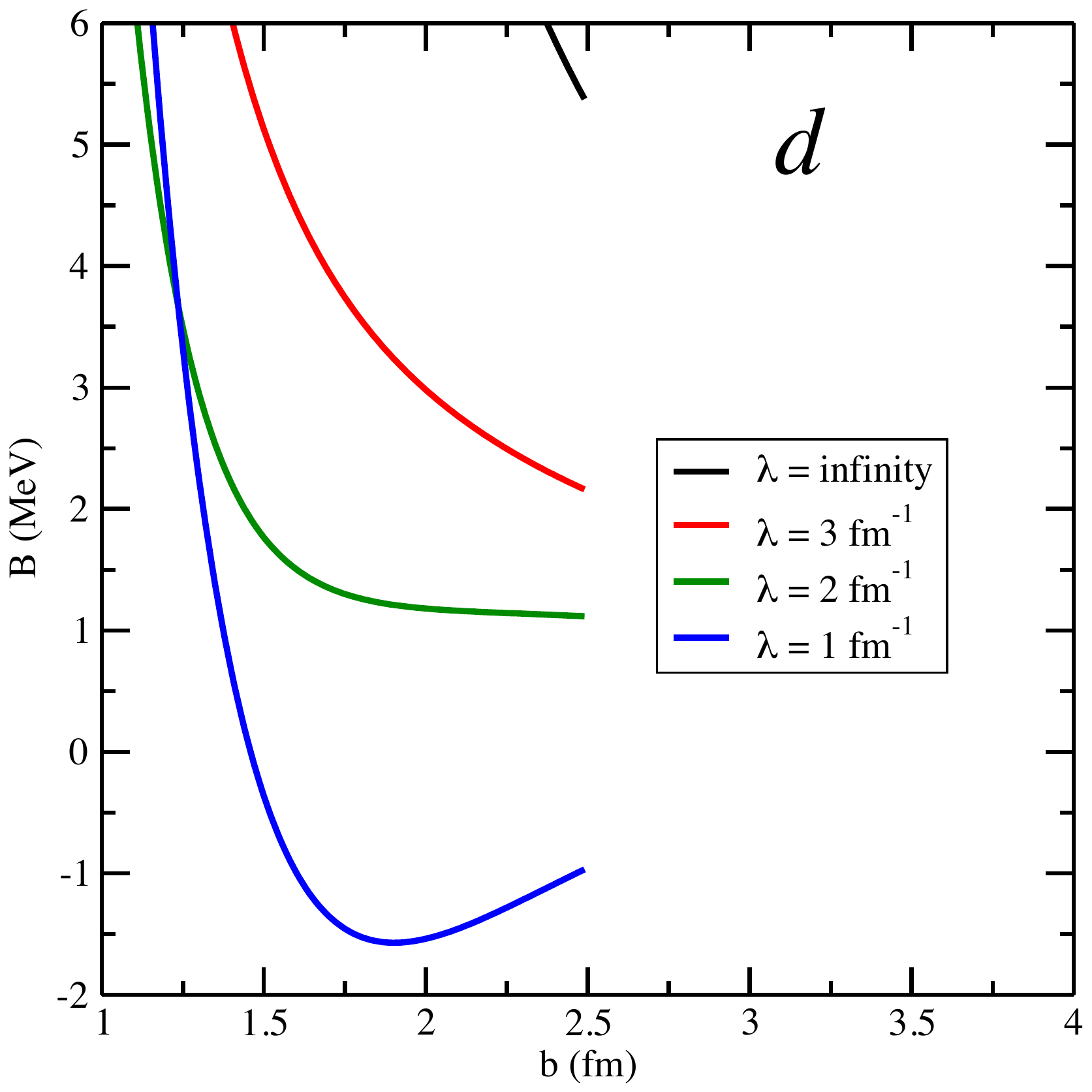}
\includegraphics[height=3.4cm,width=3.5cm]{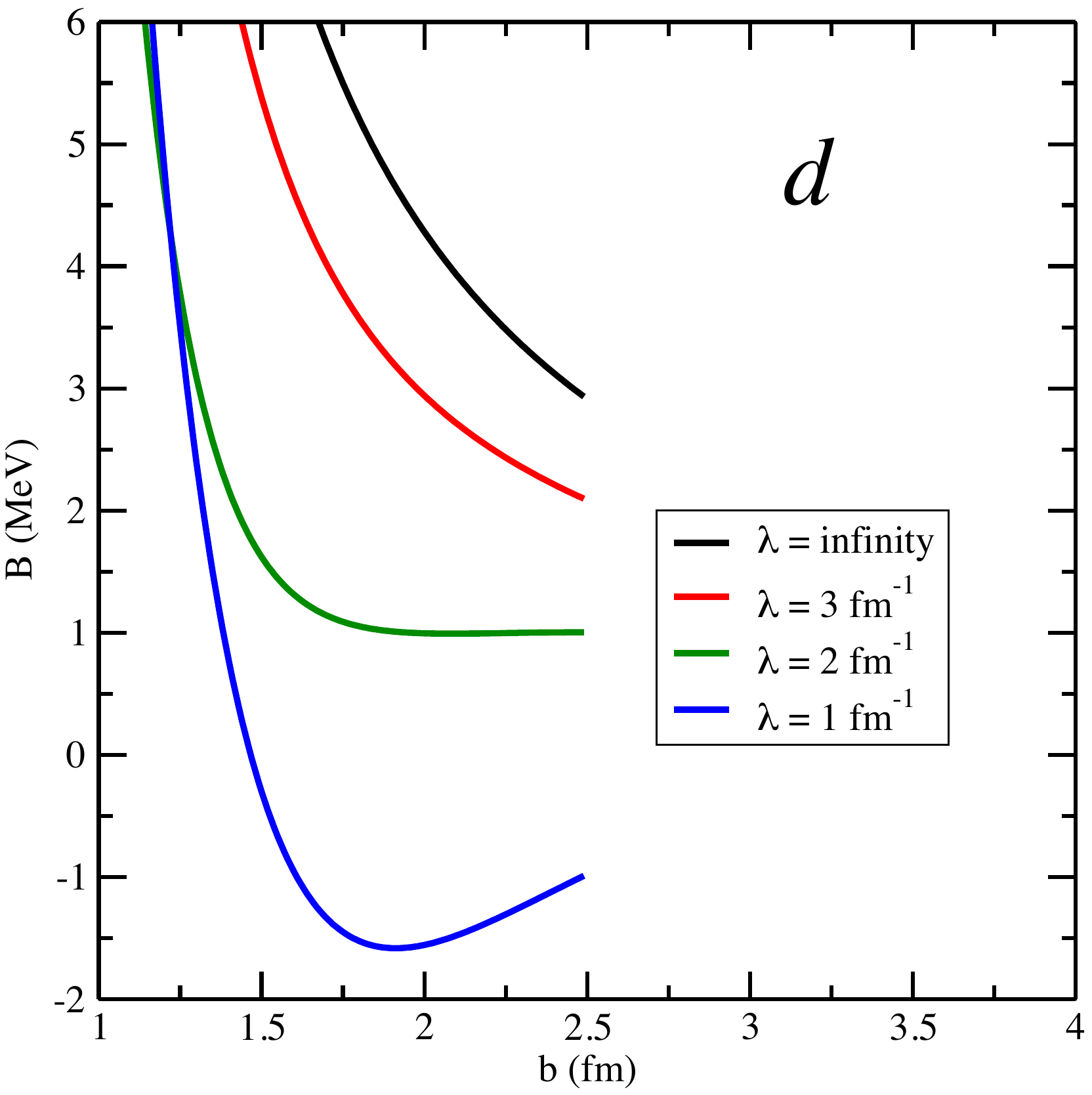}
\includegraphics[height=3.4cm,width=3.5cm]{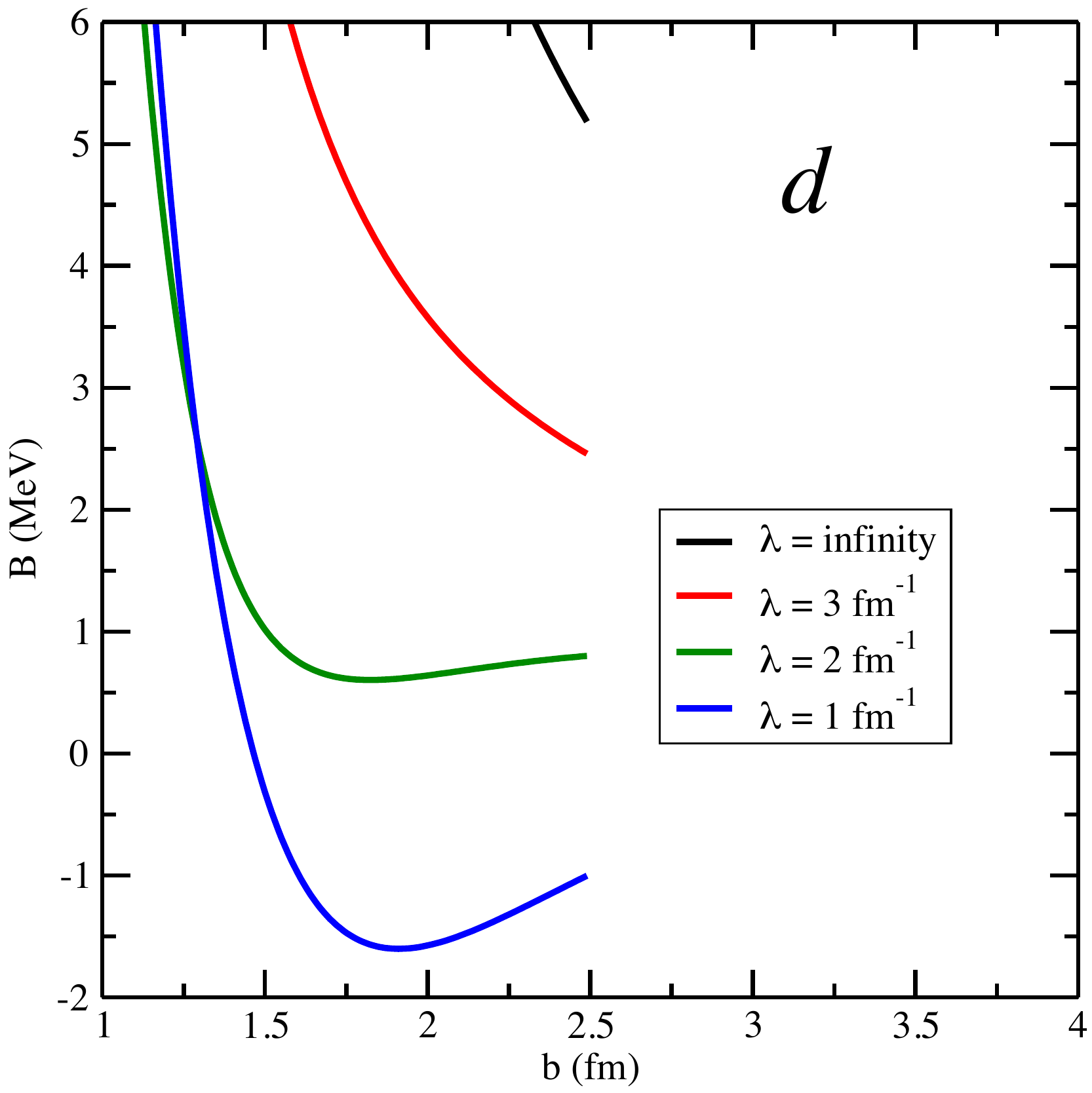} \\
\includegraphics[height=3.4cm,width=3.5cm]{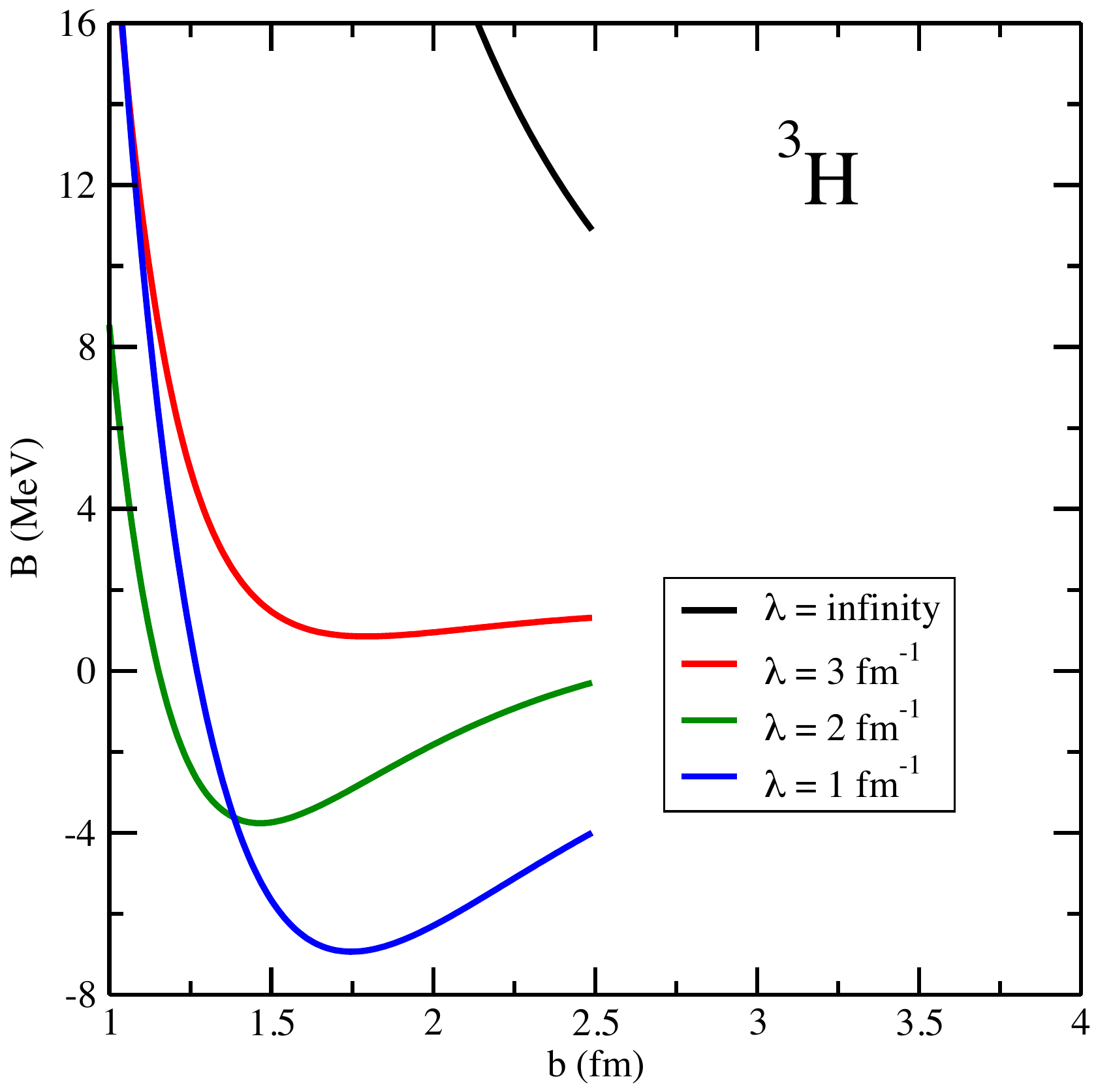}
\includegraphics[height=3.4cm,width=3.5cm]{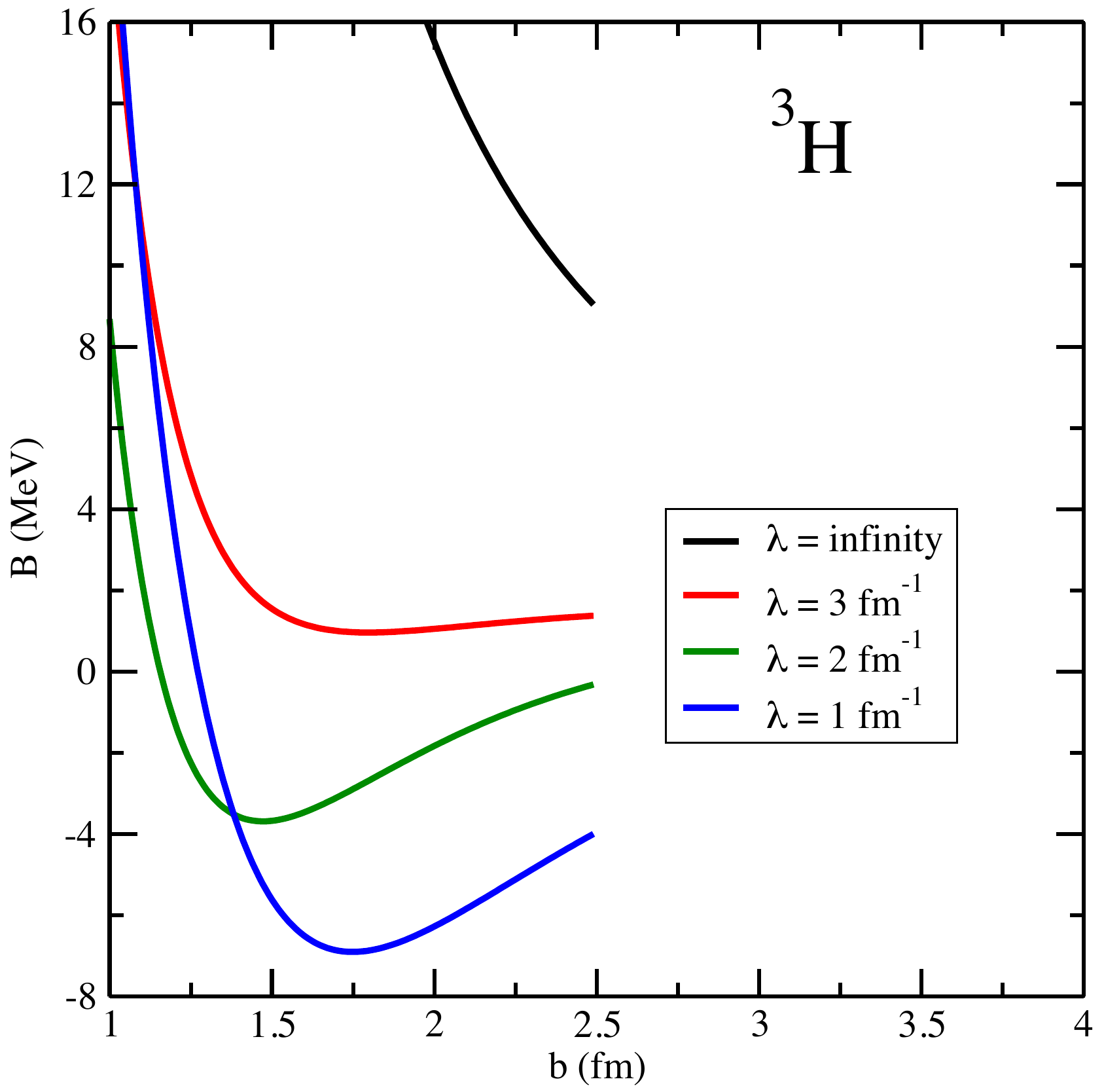}
\includegraphics[height=3.4cm,width=3.5cm]{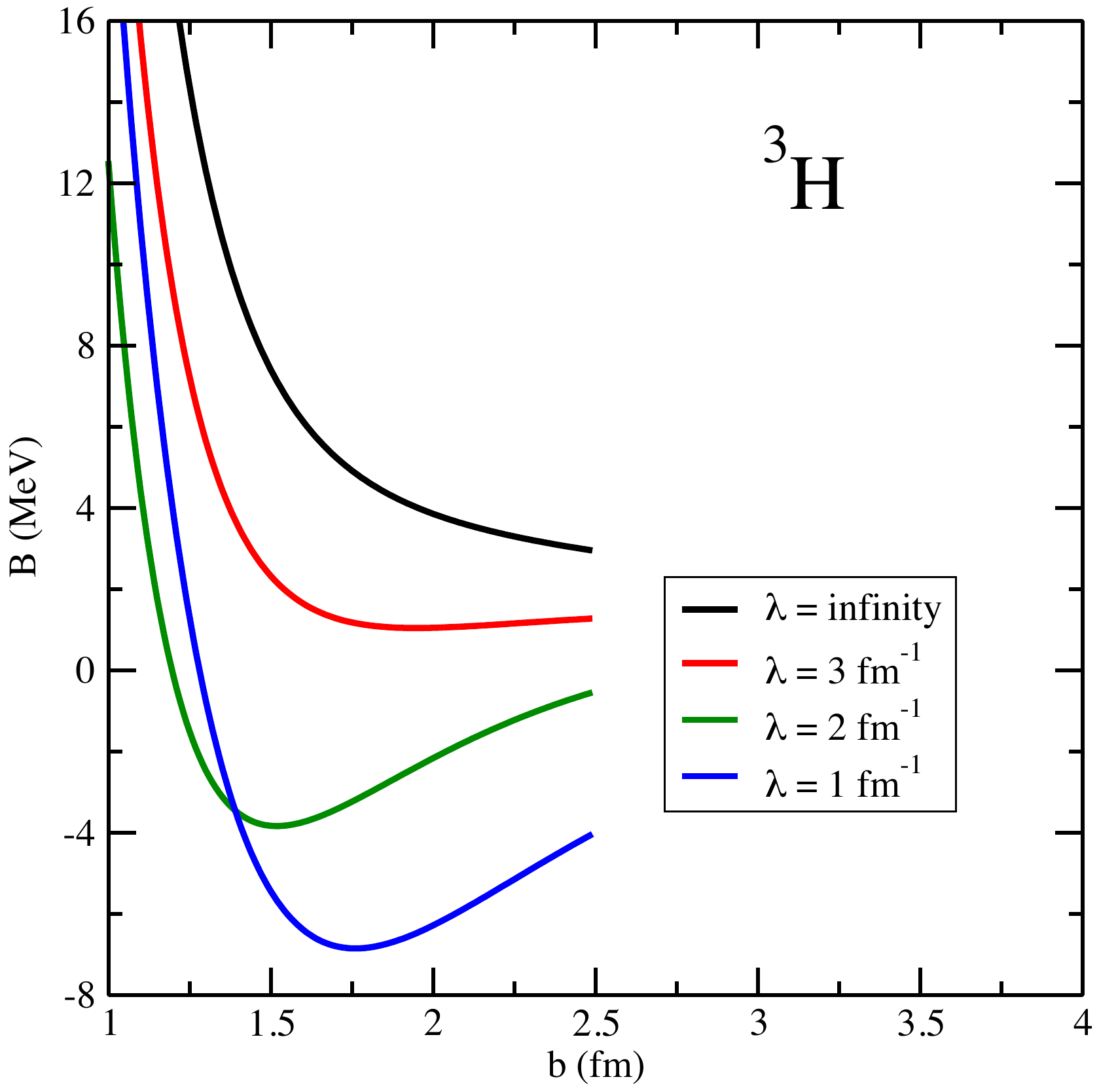}
\includegraphics[height=3.4cm,width=3.5cm]{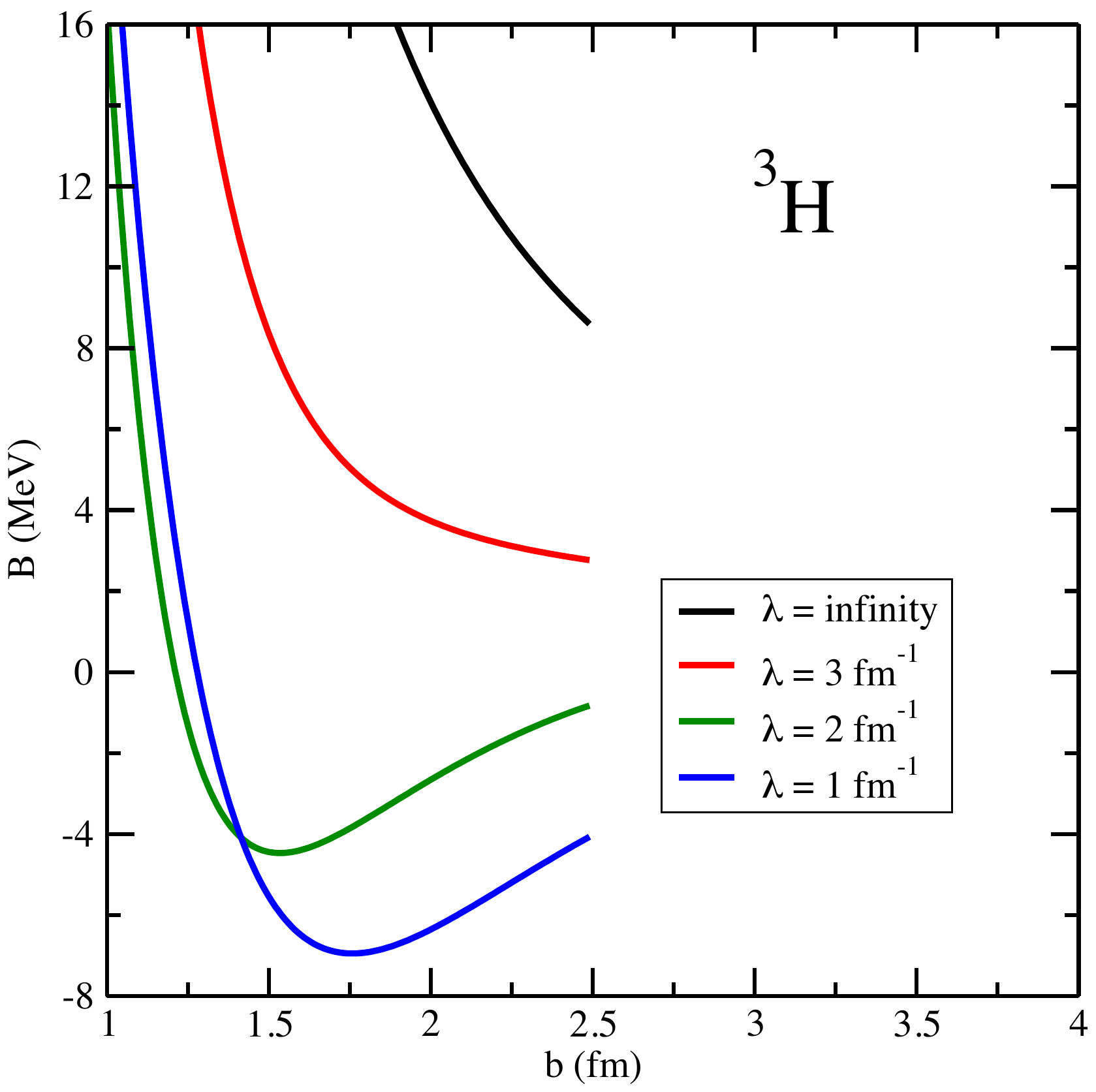} \\
\includegraphics[height=3.4cm,width=3.5cm]{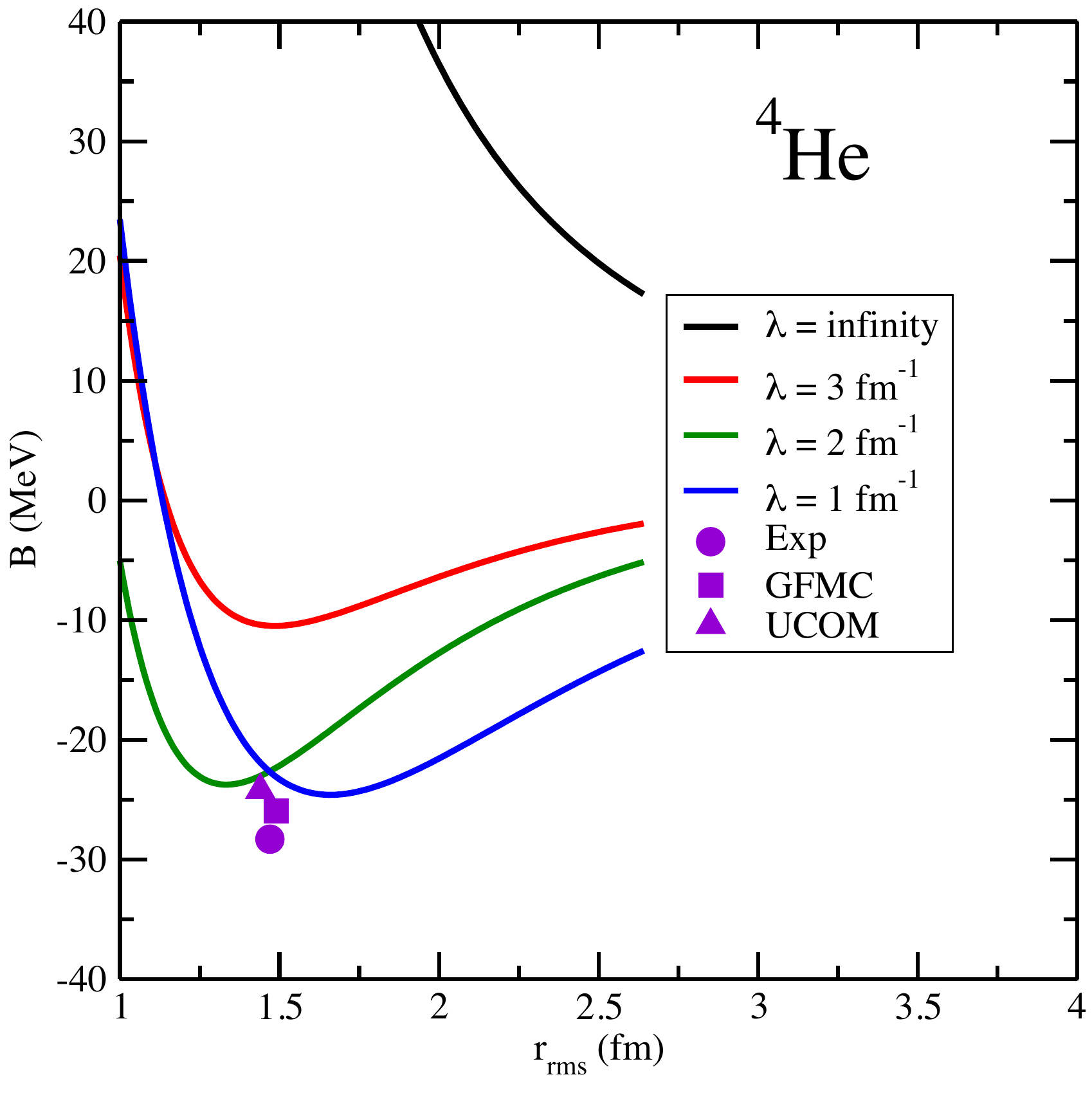} 
\includegraphics[height=3.4cm,width=3.5cm]{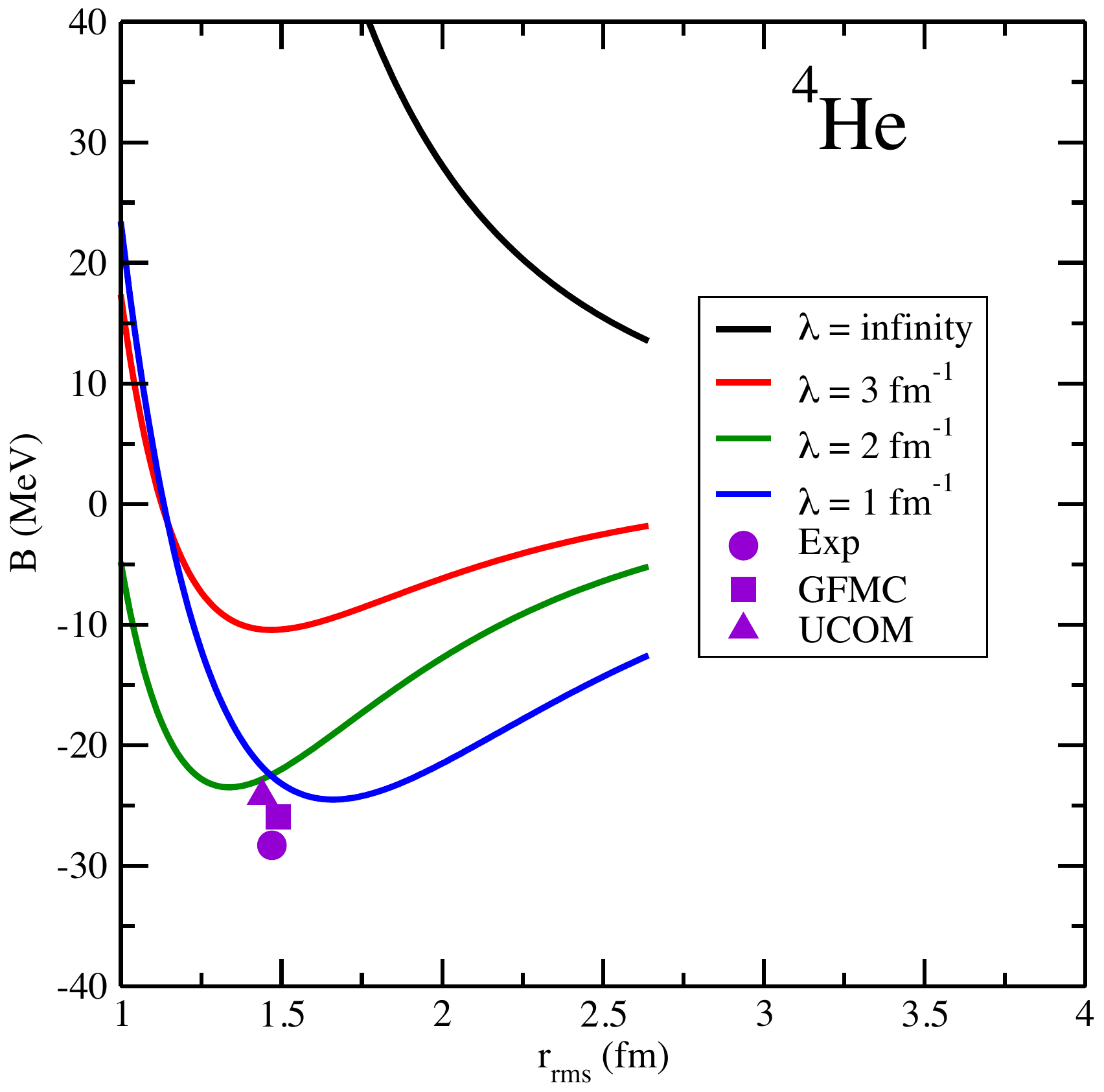} 
\includegraphics[height=3.4cm,width=3.5cm]{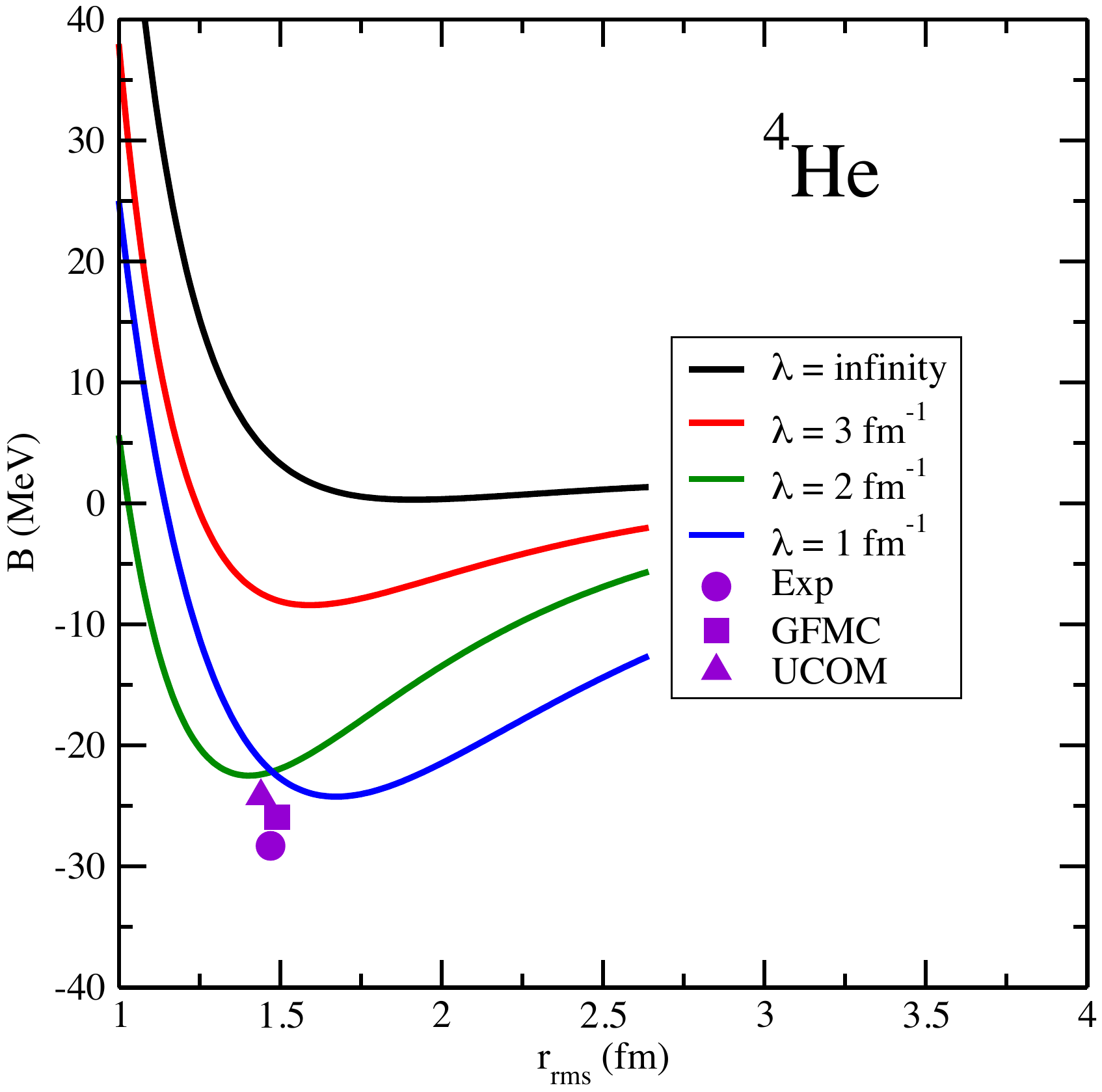} 
\includegraphics[height=3.4cm,width=3.5cm]{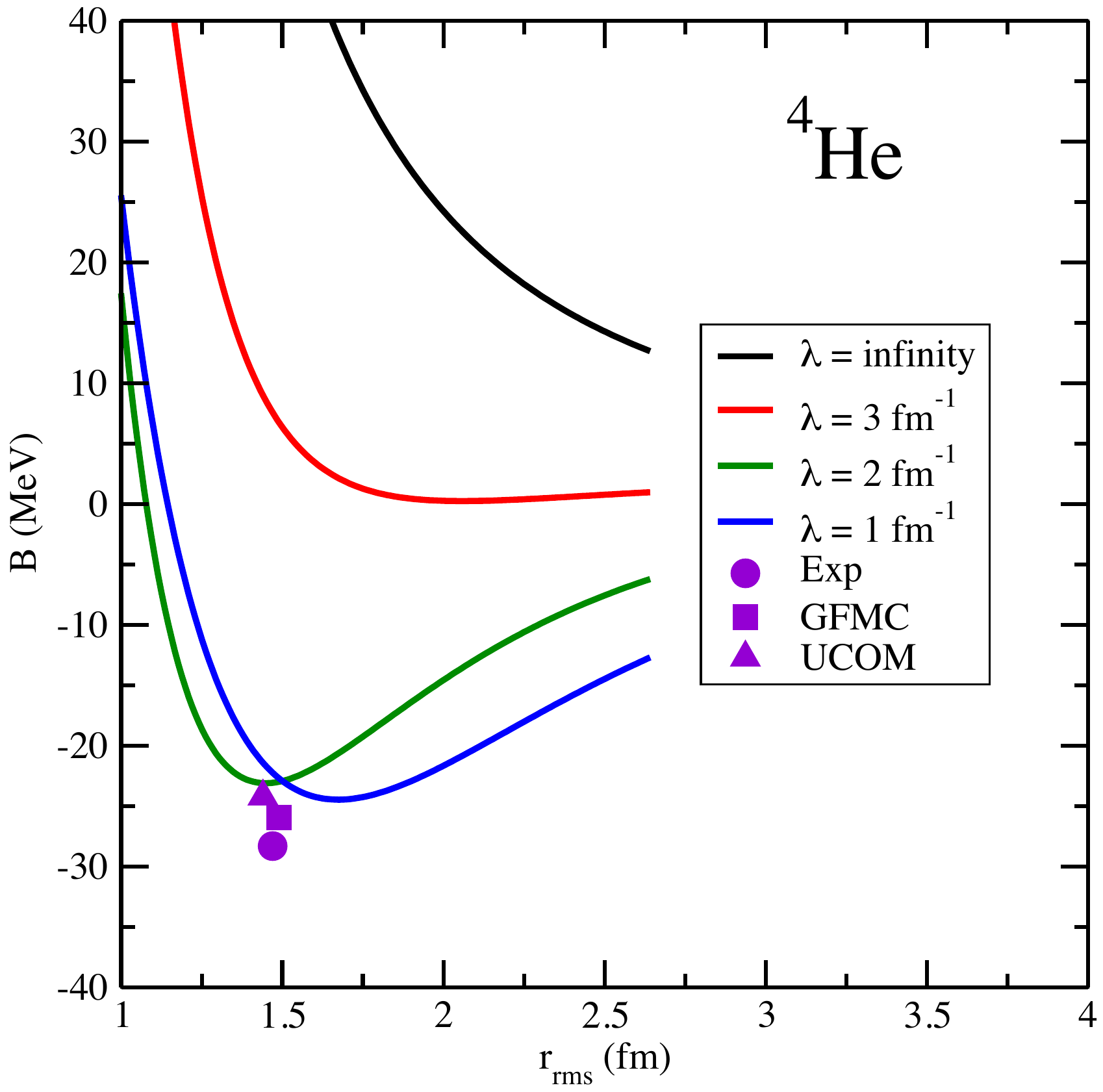} \\
\includegraphics[height=3.4cm,width=3.5cm]{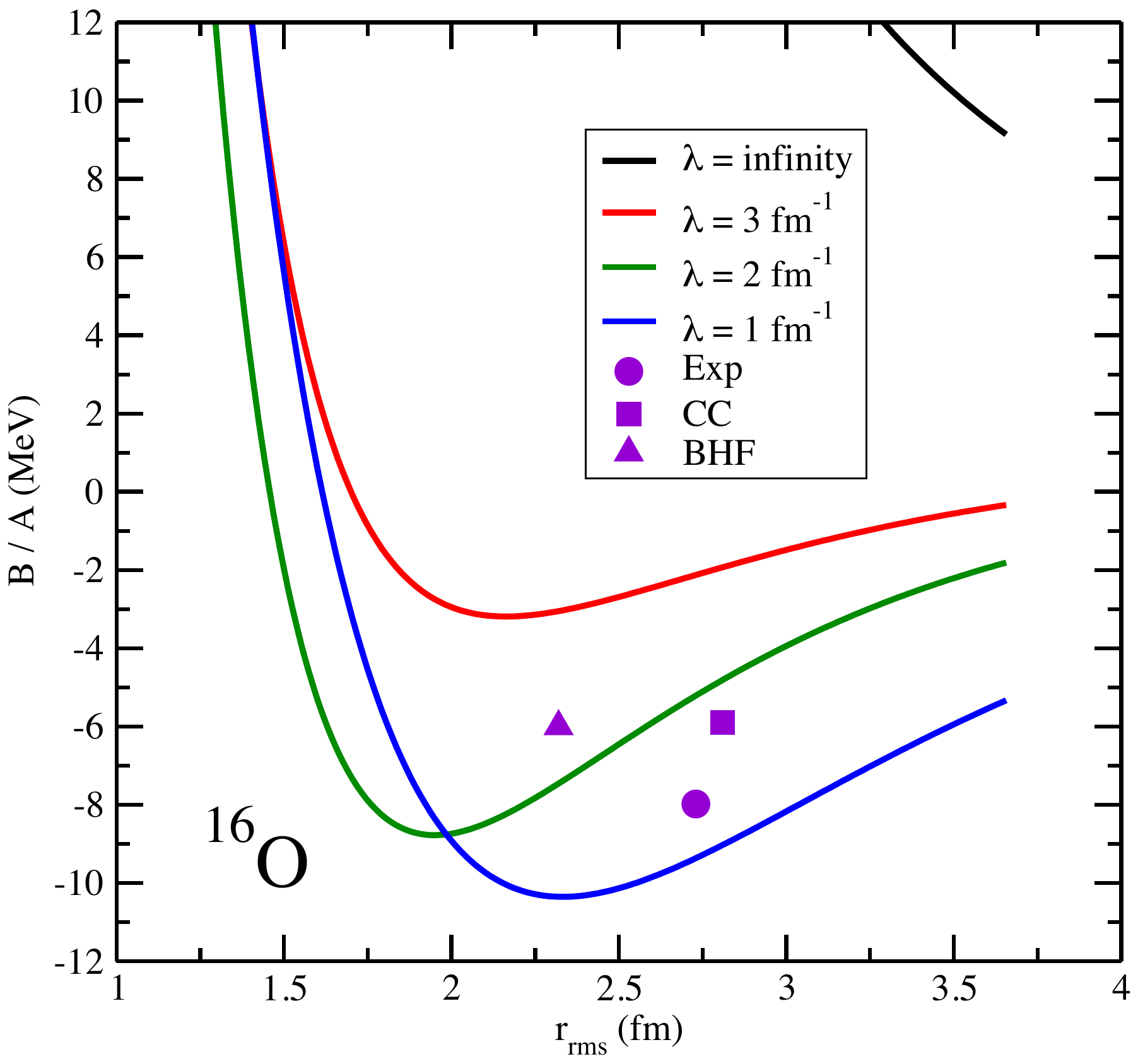}
\includegraphics[height=3.4cm,width=3.5cm]{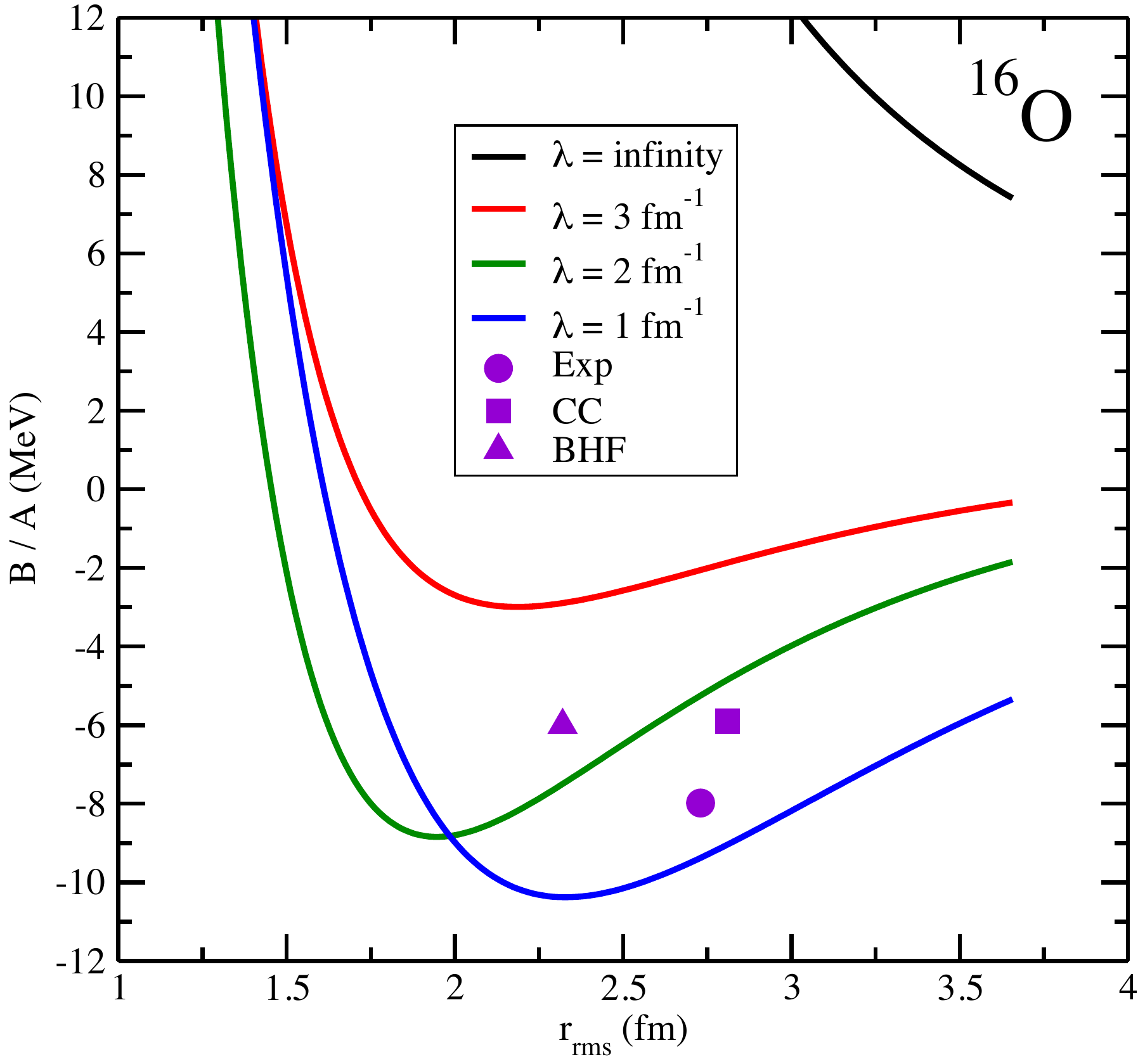}
\includegraphics[height=3.4cm,width=3.5cm]{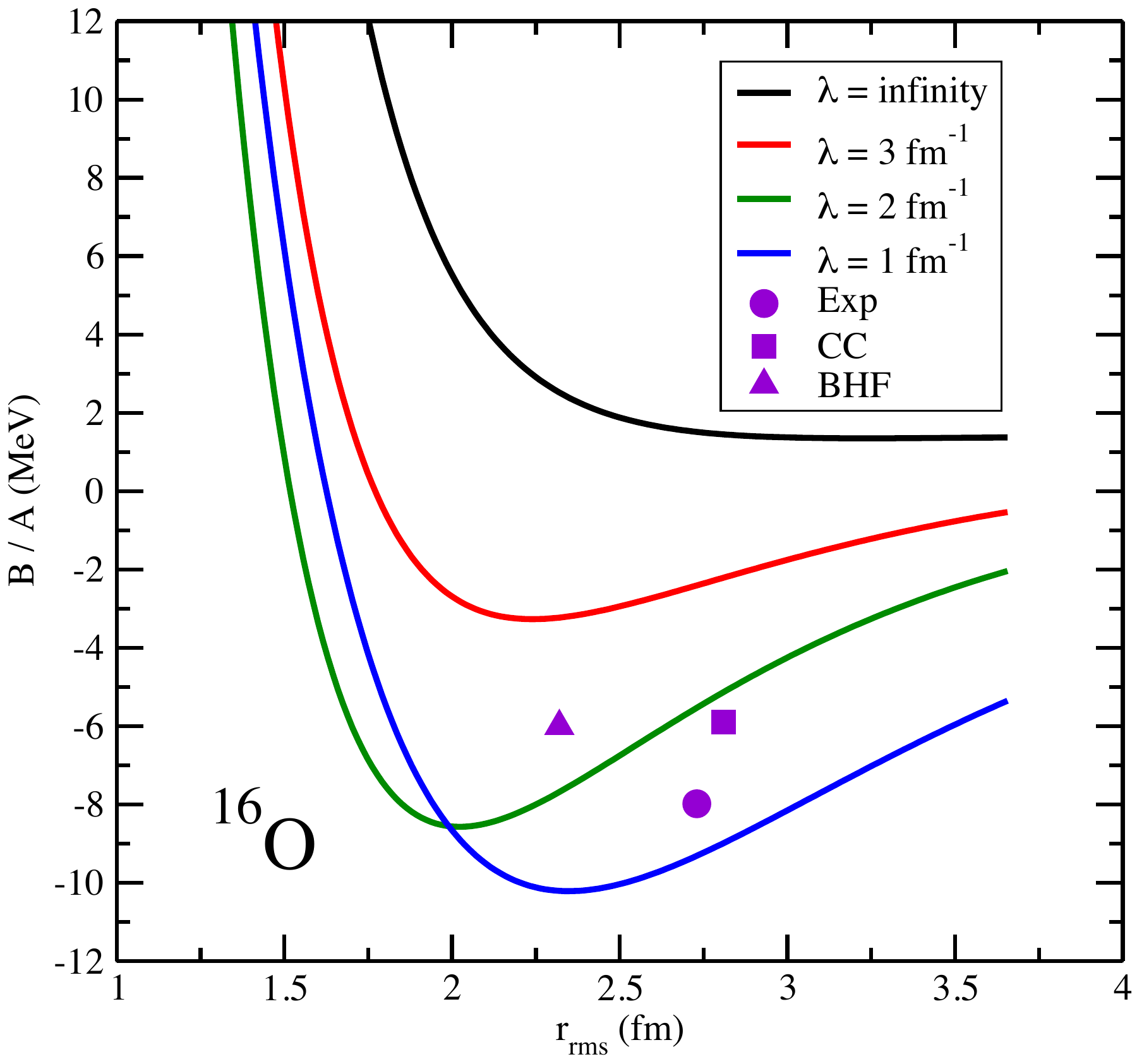}
\includegraphics[height=3.4cm,width=3.5cm]{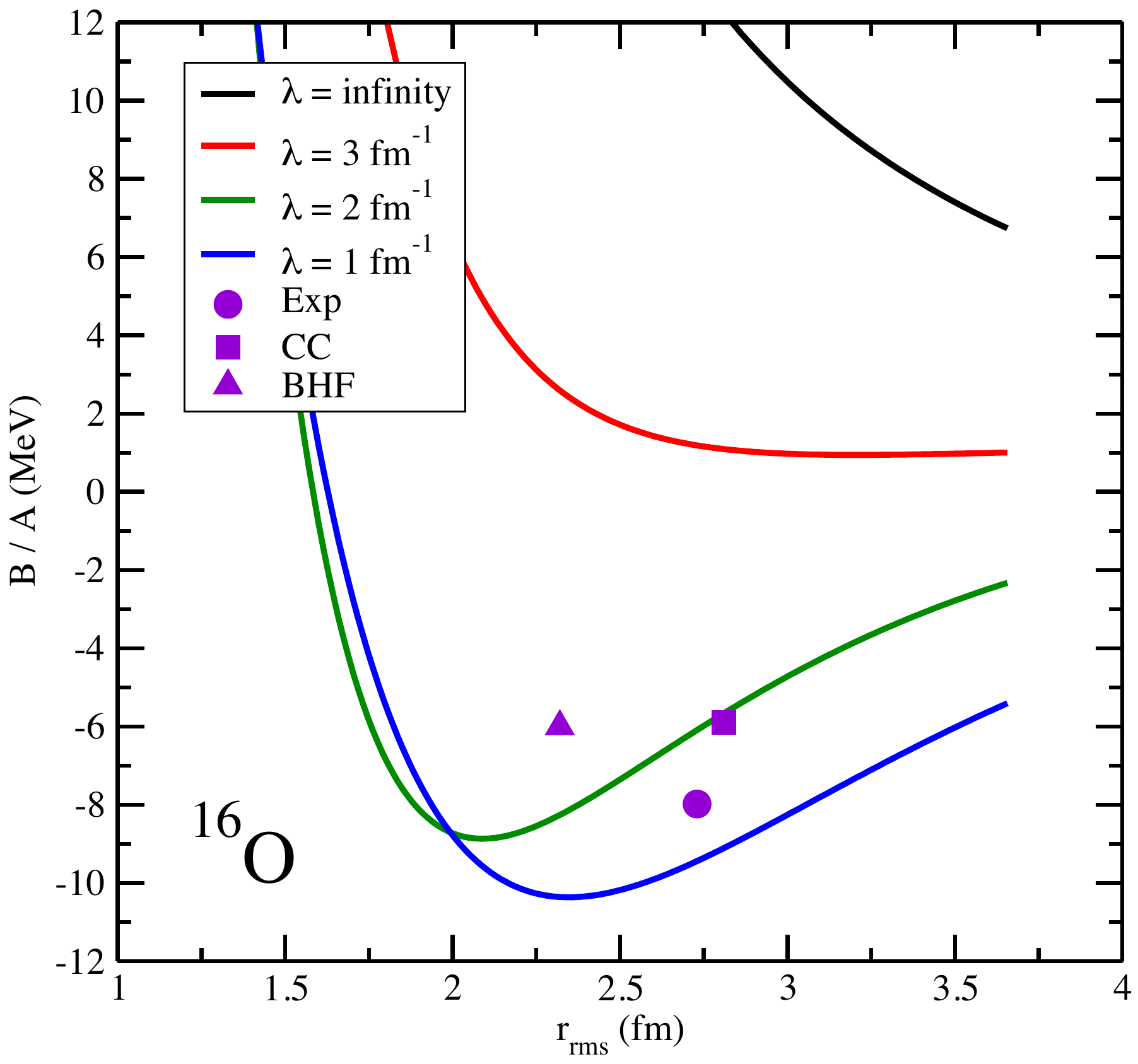} \\
\includegraphics[height=3.4cm,width=3.5cm]{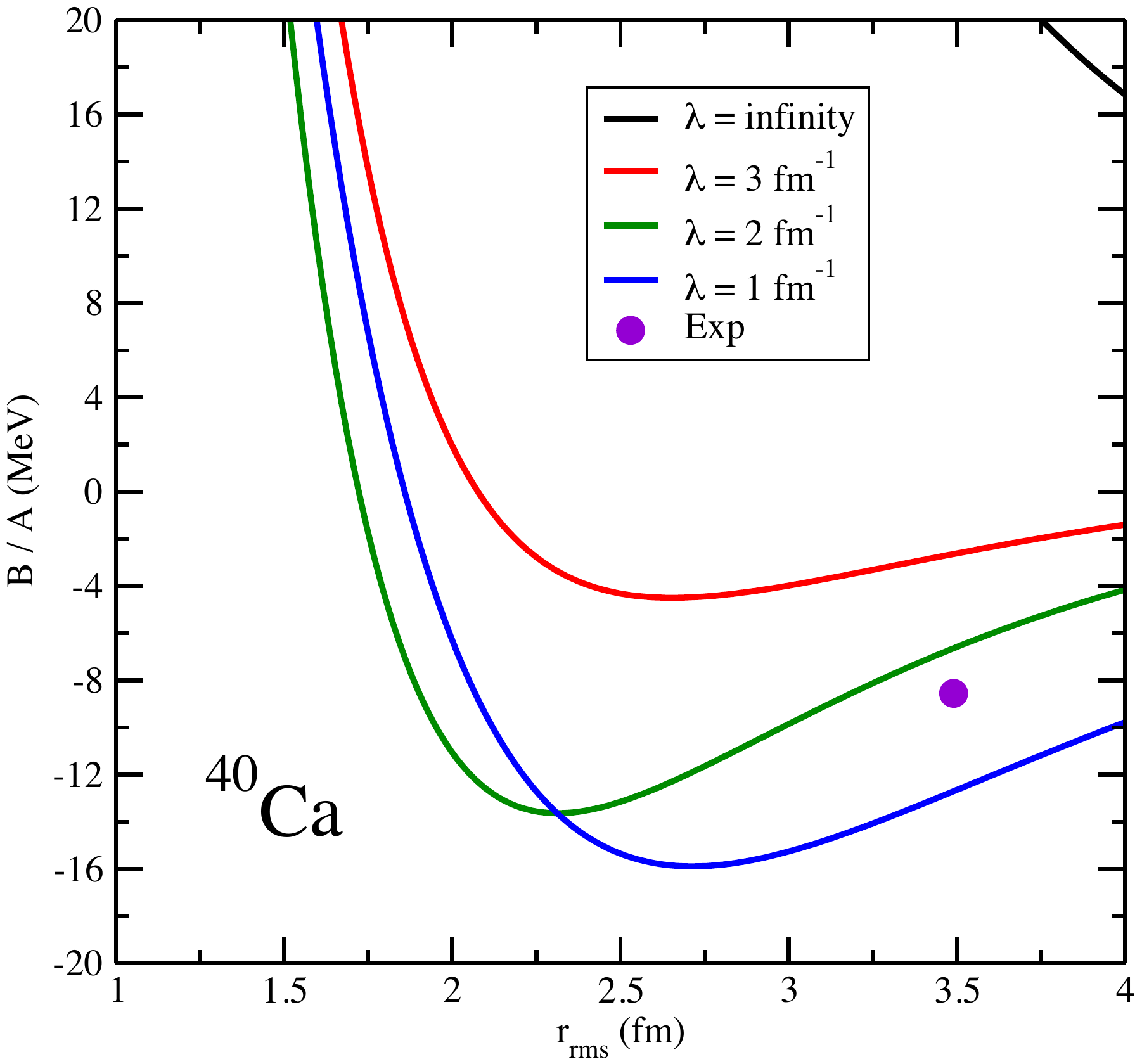}
\includegraphics[height=3.4cm,width=3.5cm]{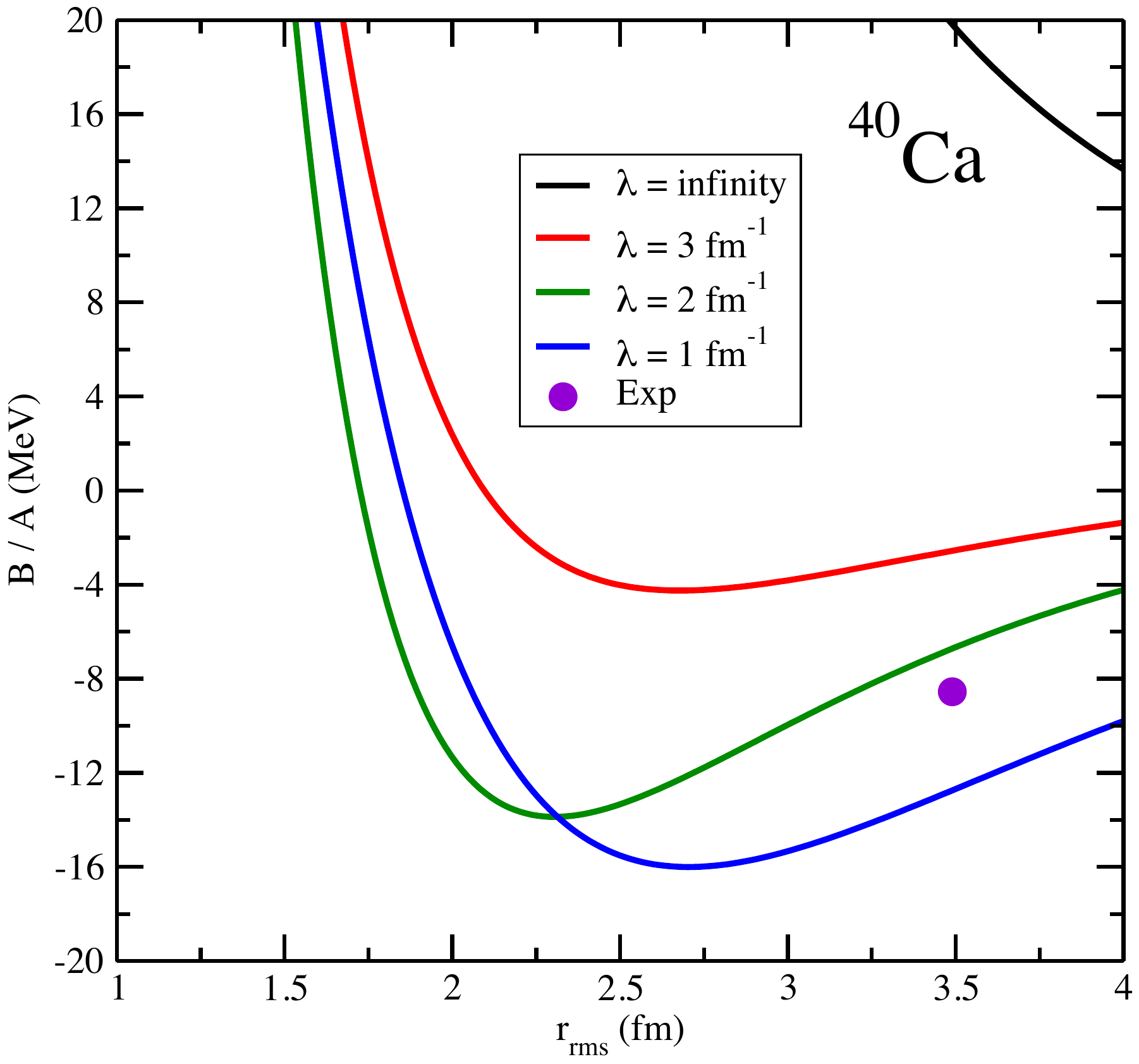}
\includegraphics[height=3.4cm,width=3.5cm]{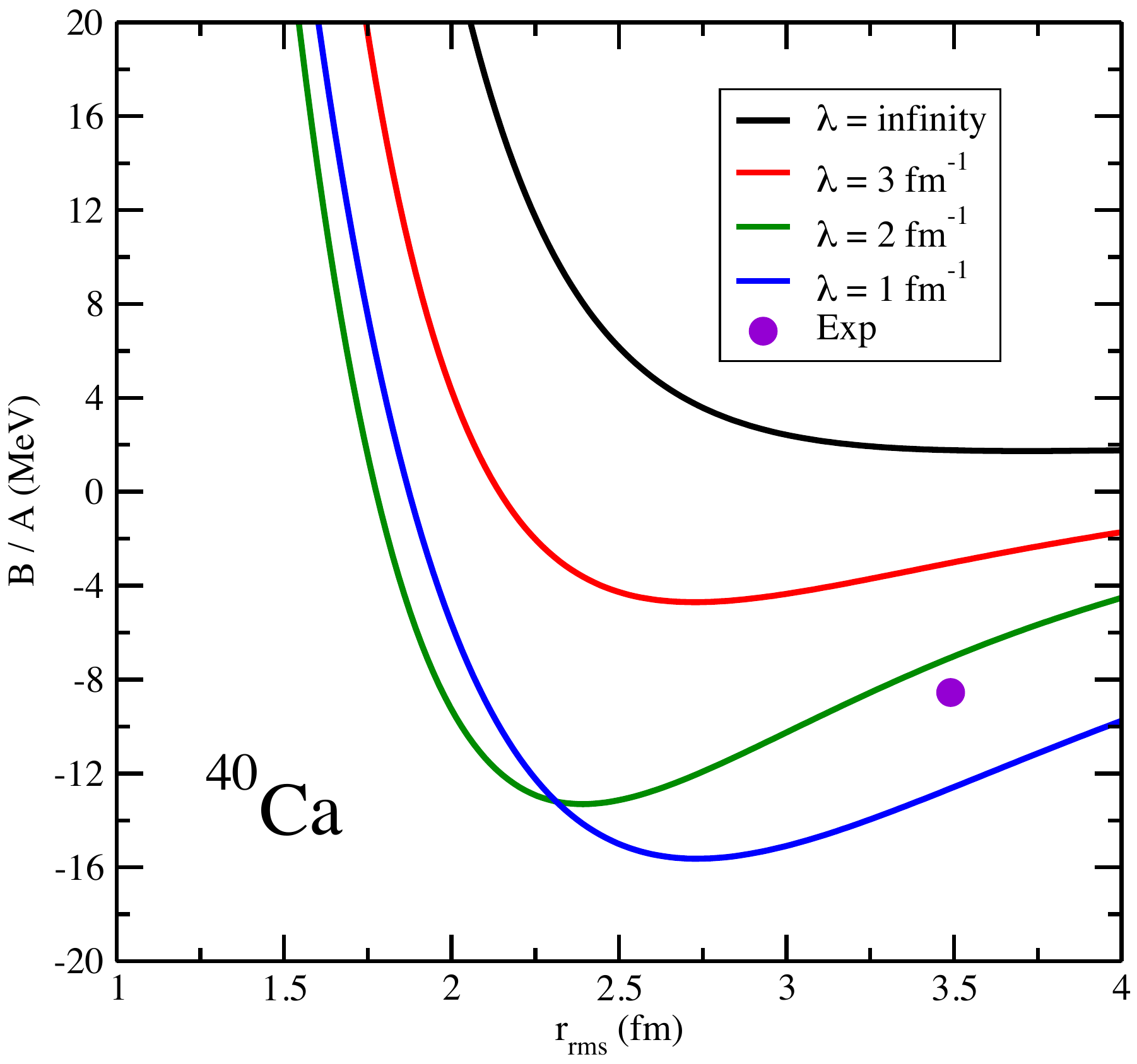}
\includegraphics[height=3.4cm,width=3.5cm]{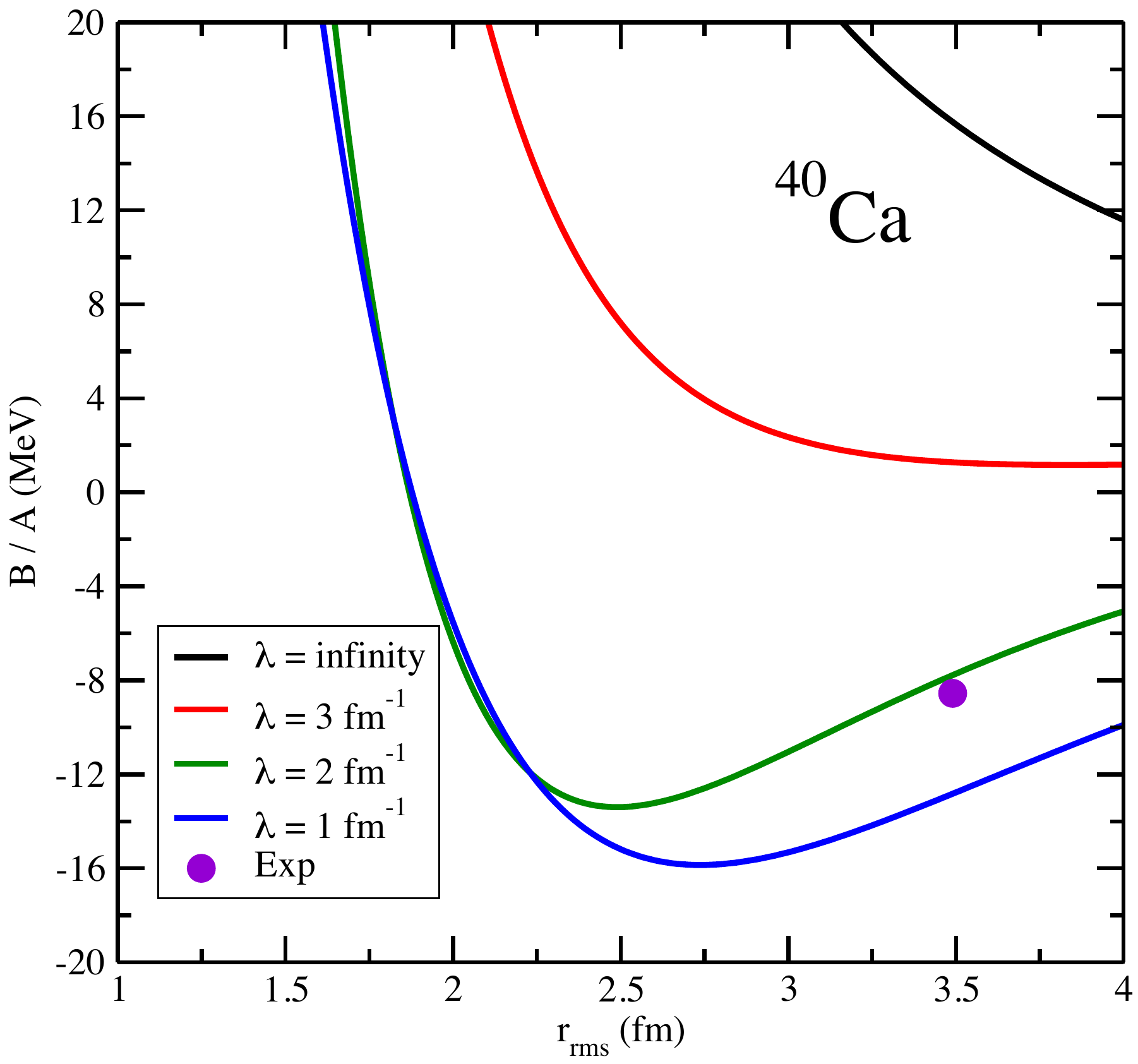} \\
\includegraphics[height=3.4cm,width=3.5cm]{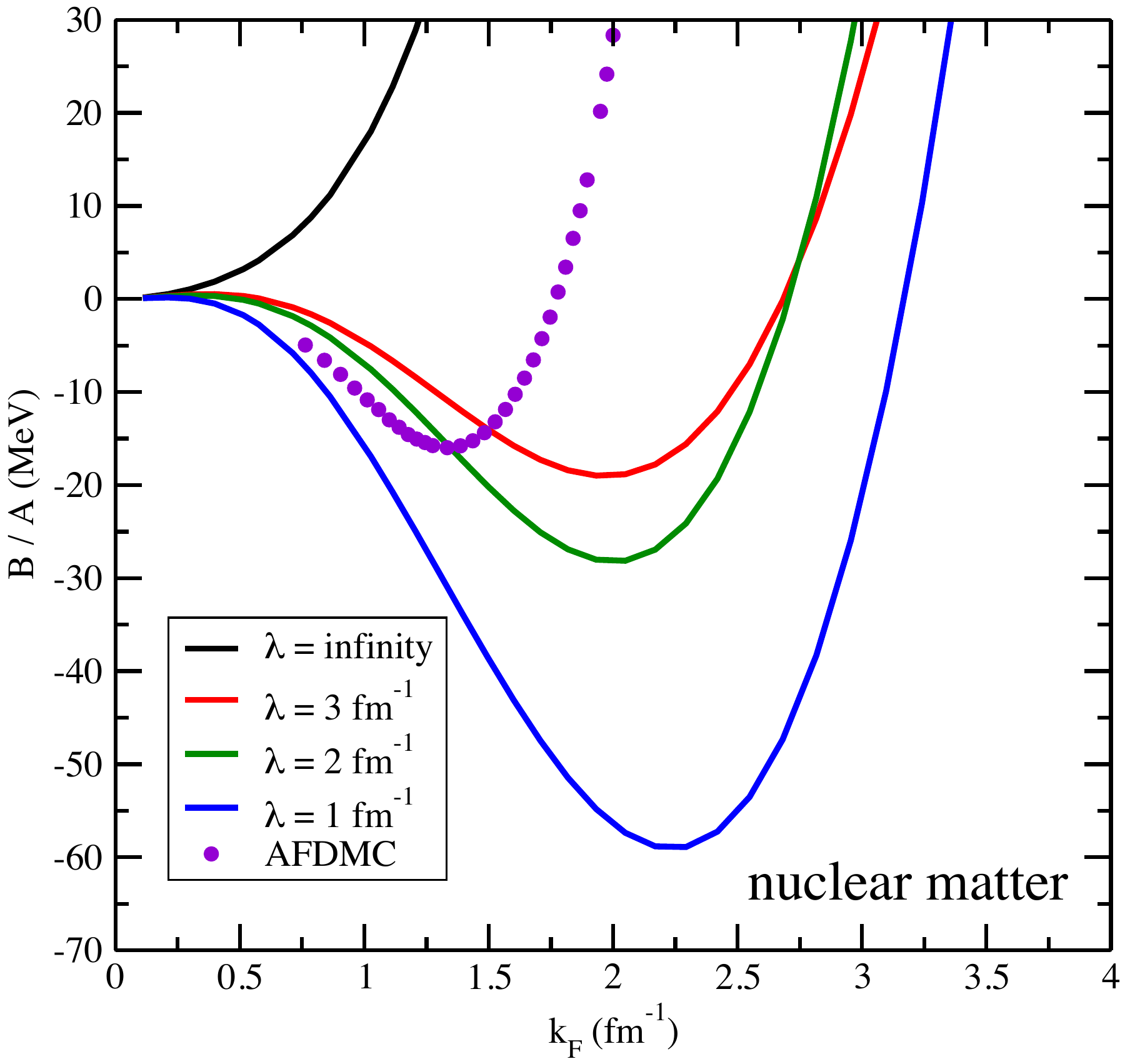} 
\includegraphics[height=3.4cm,width=3.5cm]{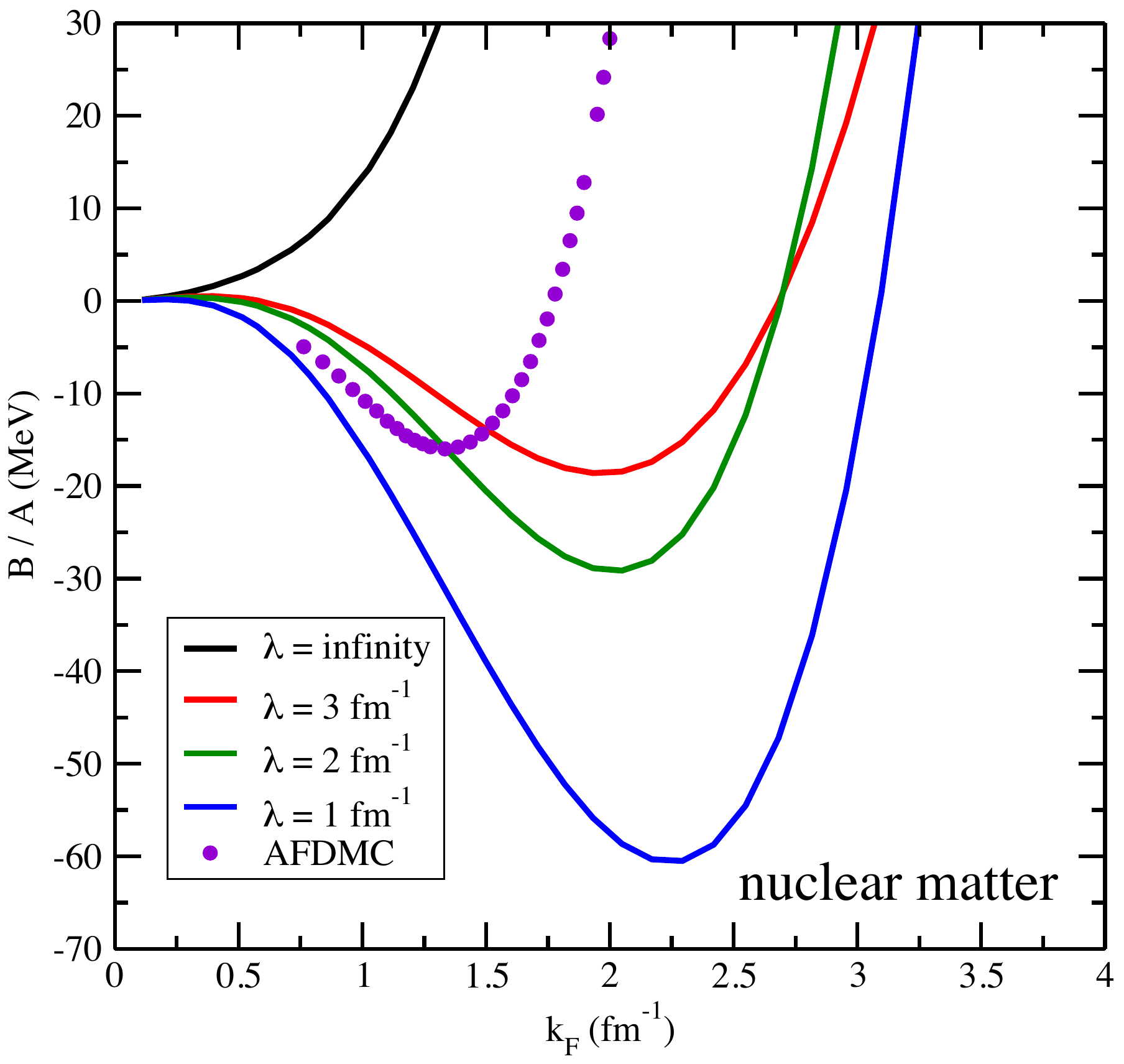} 
\includegraphics[height=3.4cm,width=3.5cm]{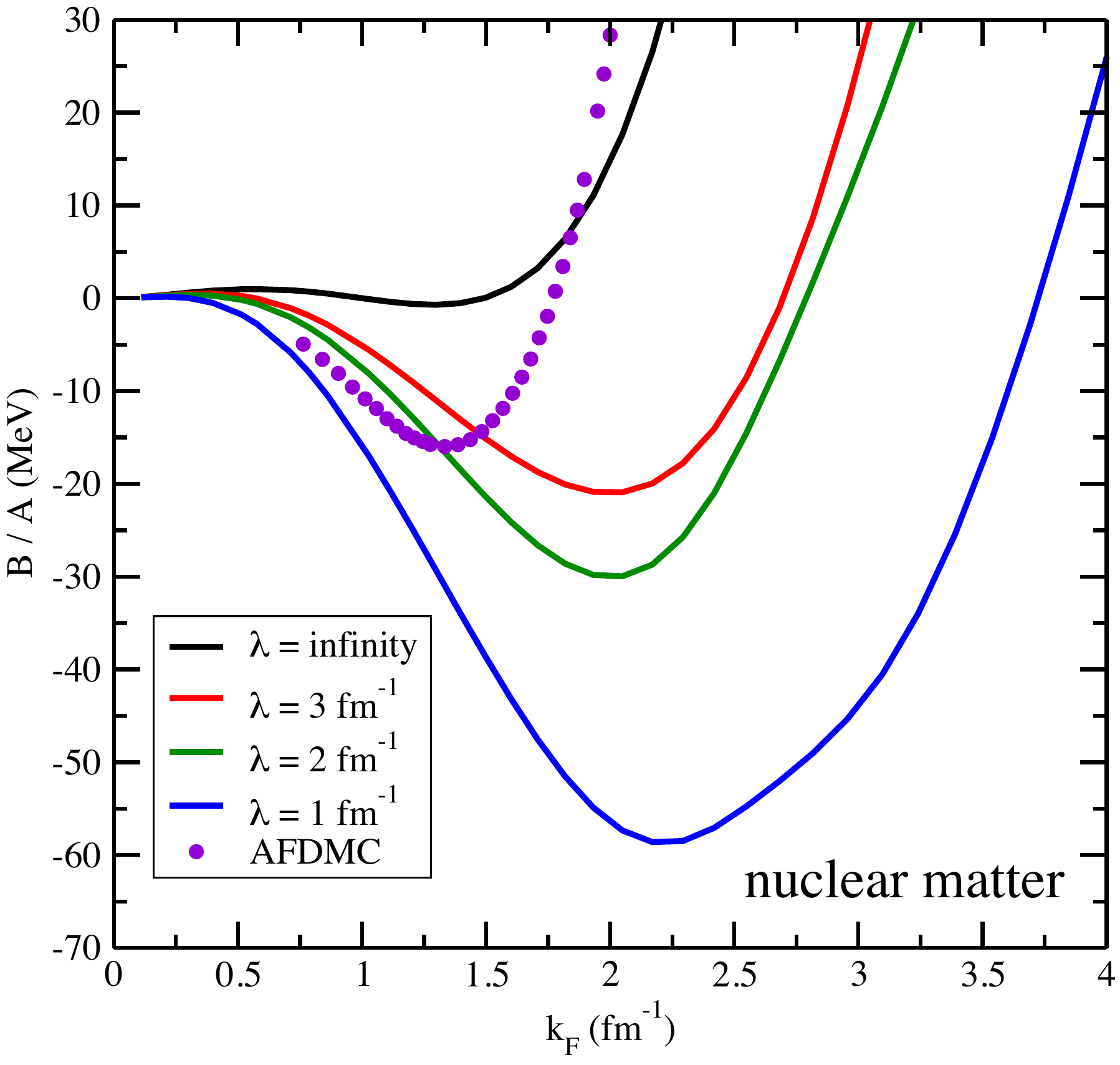} 
\includegraphics[height=3.4cm,width=3.5cm]{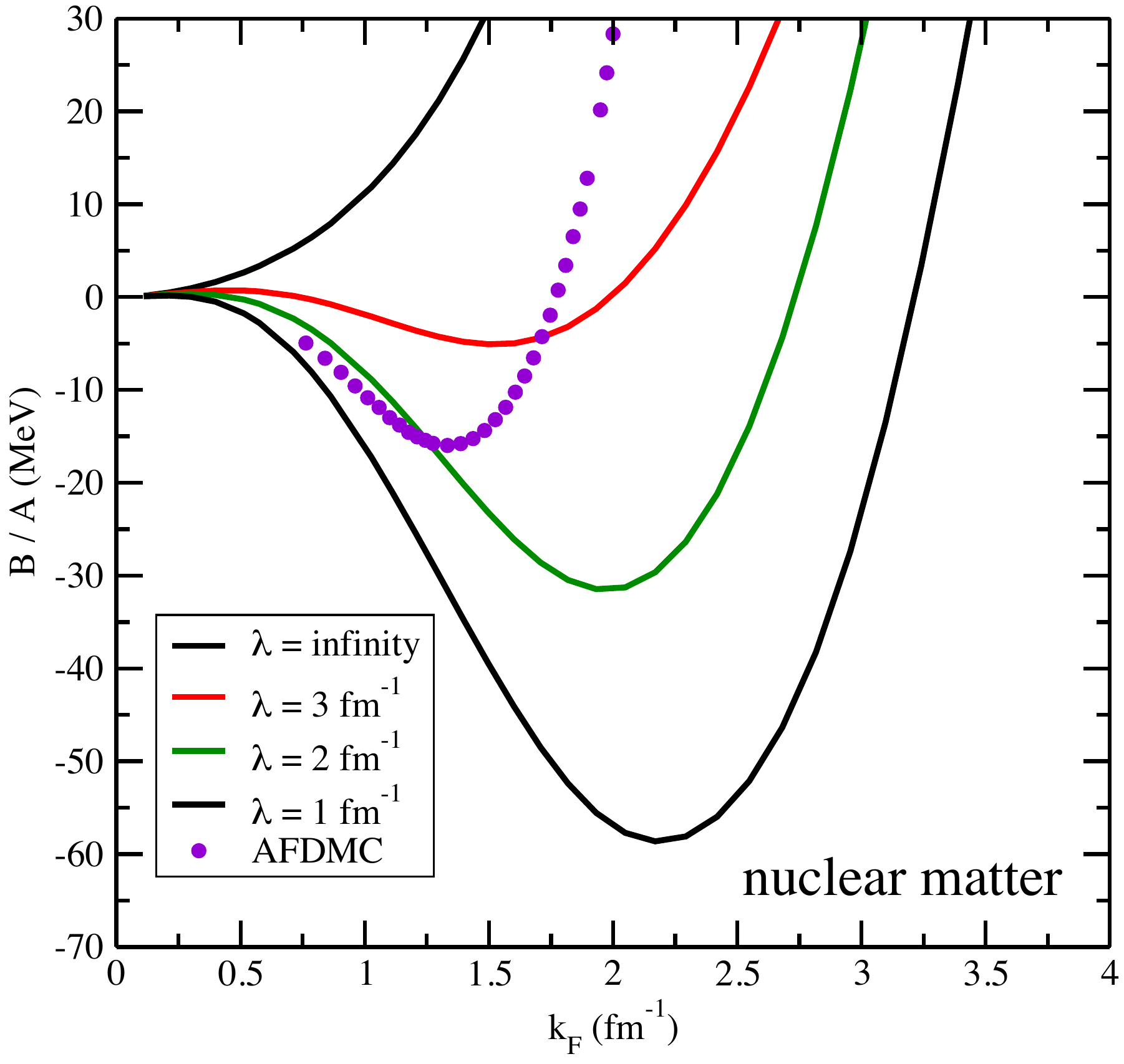}
\end{center}
\caption{Binding energies including Coulomb (in MeV) vs msr $r_{m}$
  (in fm) for $^2{\rm H}=d$, $^3{\rm H}$, $^4{\rm He}$, $^{16}{\rm O}$
  and $^{40}{\rm Ca}$ and nuclear matter for different SRG-$\lambda$
  for NijmII~\cite{Stoks:1994wp}, AV18~\cite{Wiringa:1994wb},
  N3LO-EM~\cite{Entem:2003ft} and N3LO-EGM~\cite{Epelbaum:2004fk}
  potentials (from left to right) and Harmonic Oscillator wave
  functions. We compare to some calculations and also to experimental
  data. Nuclear matter is computed in the Hartree-Fock approximation.}
\label{fig:Binding-SRG}
\end{figure}

\section{Finite Nuclei and Nuclear matter}

If is tempting to make simple estimates based on harmonic oscillator
(HO) shell model wave functions. Of course, once we make a unitary
transformation such as SRG on the two-body sector we are effectively
generating multinucleon forces~\cite{Jurgenson:2009qs}. From a
practical point of view, 3- and 4-nucleon forces are so far fixed from
$^3{\rm H}$ or $^4{\rm He}$ binding energies, and they turn out to
almost vanish at about $\lambda \sim 2 {\rm
  fm}^{-1}$~\cite{Jurgenson:2009qs}.  From Fig.~\ref{fig:Binding-SRG}
we see that for $\lambda \sim 1 {\rm fm}^{-1}$ we get $( B_d^{\rm HO},
B_{\rm 3H}^{\rm HO}, B_{\rm 4He}^{\rm HO} )= ( -1.5 , -7.1,-24.0) {\rm
  MeV} $, close (except for $B_d$) to more accurate
calculations~\cite{Jurgenson:2009qs}, regardless on using
NijmII~\cite{Stoks:1994wp}, AV18~\cite{Wiringa:1994wb} and the chiral
N3LO-EM~\cite{Entem:2003ft} and N3LO-EGM~\cite{Epelbaum:2004fk}
potentials. This is in line with the coarse grained potentials
calculation~\cite{NavarroPerez:2011fm} and
%\begin{eqnarray}
%{\rm min}_{\Psi}\langle \Psi | H_{\rm Av18}^{\lambda=\infty}|
%\Psi\rangle \sim {\rm min}_b \langle (1s)^4 | H_{\rm Av18}^{\lambda
%  \sim 1{\rm fm}^{-1}} | (1s)^4\rangle \ \sim {\rm min}_b \langle
%(1s)^4 | H_{\rm N3LO}^{\lambda \sim 1{\rm fm}^{-1}} | (1s)^4\rangle
%\end{eqnarray} suggests that both the AV18 core and the (Chiral) TPE
are marginal, at least for the $^3$H and $^4$He binding. Further,
working at the SRG Wigner scale $\lambda_{\rm Wigner}= 3 {\rm
  fm}^{-1}$ gives unbound triton and a poor value $B_{\rm 4He}^{\rm
  HO} = -10 {\rm MeV} $, while in Ref.~\cite{Jurgenson:2009qs} it is
found $(B_{\rm 3H}, B_{\rm 4He})= (-8.1,-26.8) {\rm MeV} $.  Nuclear
matter in the Hartree-Fock approximation saturates, although it
describes a Coester-like band along the SRG trajectory typical of
two-body interacions.  This of course raises the question on whether
or not the needed 3- and 4-body forces are
SU(4)-invariant (see e.g.\cite{Birse:2012ih}).

\section{Conclusions}

From a fundamental viewpoint, QCD large $N_c$ based arguments foresee
fulfilling Wigner symmetry with a relative ${\cal O}(N_c^{-2})$
accuracy.  This complies with our finding at $\lambda_{\rm Wigner}$, a
remarkable and surprissing result if we consider that nowhere in the
design and optimization of the modern high-quality interactions was
the Wigner symmetry pattern explicitly implemented.

%\bibliographystyle{unsrt}
%\bibliography{refs}

\end{document}